\documentclass[pageno]{jpaper}

\usepackage[normalem]{ulem}
\usepackage{cite}
\usepackage{amsmath,amssymb,amsfonts}
\usepackage{algorithmic}
\usepackage{graphicx}
\usepackage{textcomp}
\usepackage{xcolor}
\usepackage{fancyhdr}

\usepackage[normalem]{ulem}

\usepackage{soul}

\usepackage{booktabs}
\usepackage{multirow}
\usepackage{xspace}
\usepackage{flushend}

\newcommand{\ignore}[1]{}
\newcommand{\todo}[1]{\textcolor{red}{\sf\bfseries #1}}
\newcommand{\hwk}{\textsf{HawkEye}}
\newcommand{\trident}{\textsf{Trident}}
\newcommand{\smartcompaction}{\textsf{smart compaction}}
\newcommand{\tridentG}{\textsf{Trident1G}-\textsf{only}}
\newcommand{\thp}{\textsf{\small THP}}
\newcommand{\newtext}[1]{\textcolor{blue}{ #1}}
\newcommand{\tridentpv}{\textsf{Trident\textsuperscript{pv}}}

\newcommand{\highlight}[1]{#1}
\newcommand{\circled}[1]{\textcircled{\textsf{#1}}}

\newcommand{\gva}{\textsf{gVA}}
\newcommand{\gpa}{\textsf{gPA}}

\newcommand{\vs}[1]{\vspace{#1}}
\renewcommand{\vs}[1]{}

\begin{document}

\title{\LARGE Leveraging Architectural Support of Three Page Sizes with Trident}
\author{
\Large {Venkat Sri Sai Ram, Ashish Panwar, Arkaprava Basu} \\
$\{$sirvisettis, ashishpanwar, arkapravab$\}$@iisc.ac.in \\
%Computer Systems Lab (\url{https://csl.csa.iisc.ac.in/}) \\
Department of Computer Science and Automation \\
Indian Institute of Science
}

\date{}
\maketitle

\thispagestyle{empty}

\begin{abstract}
Large pages are commonly deployed to reduce address translation overheads for big-memory workloads. Modern x86-64 processors from Intel and AMD support two large page sizes -- 1GB and 2MB. However, previous works on large pages have primarily focused on 2MB pages, partly due to lack of substantial evidence on the profitability of 1GB pages to real-world applications. We argue that in fact, inadequate system software support is responsible for a decade of underutilized hardware support for 1GB pages.

Through extensive experimentation on a real system, we demonstrate that 1GB pages can improve performance over 2MB pages, and when used in tandem with 2MB pages for an important set of applications; the support for the latter is crucial but missing in current systems. Our design and implementation of \trident{} in Linux fully exploits hardware supported large pages by dynamically and transparently allocating 1GB, 2MB, and 4KB pages as deemed suitable. \trident{} speeds up eight memory-intensive applications by {$18\%$}, on average, over Linux's use of 2MB pages. We also propose \tridentpv{}, an extension to \trident{} that effectively virtualizes 1GB pages via copy-less promotion and compaction in the guest OS. Overall, this paper shows that adequate software enablement can bring practical relevance to even GB-sized pages, in turn motivating architects to continue investing/innovating in large pages.
\end{abstract}

\section{Introduction}
It is not uncommon for certain architectural features to require software enablement.
%The utility of certain architectural features can only be assessed with a robust software implementation.
Unfortunately, it is also common to find those hardware features that are underutilized or mostly ignored due to the lack of adequate software support.  
This is, while paying both the runtime cost of having the new hardware, e.g., power dissipation due to the feature, and the one-time hardware design and verification cost. 
Further, architects are left in the dark about the extent to which those features are beneficial in practice, and whether they should continue enhancing them or drop them in future products. 

In this work, we shed light on one such hardware feature that has been languishing for a decade -- support for 1GB pages and provide a detailed system software (here, Linux and KVM) enablement  for the same. 

\ignore{
The virtual memory is a classic example of hardware-software co-design. 
The hardware (e.g., x86 or ARM processors) performs the virtual to physical address translation on every load/store to ensure low overhead.
The OS and/or the hypervisor (henceforth called \textit{system software}) are responsible for creating the address mappings, ensuring flexibility in memory management. 
This division of labor, however, may necessitate both hardware support and software enablement to adopt a new feature or enhancement. 
For example, the page sizes are fixed by the hardware, but a new page size is useless without system software enabling it to the software ecosystem.

The use of large pages is key to reduce address translation overheads for applications with large memory footprints~\cite{panwar:asplos:2018, panwar:asplos:2019, kwon:osdi:2016}.  
A \textit{large page} maps a larger virtual address range to contiguous physical address (e.g., 2MB or 1GB compared to default 4KB).
A single TLB (Translation Look-aside Buffer)
%~\footnote[2]{Hardware TLBs keep recently used address translations. A hit in a TLB is fast, but a miss in a TLB triggers a slow page table walk.}
entry for a large page can thus map a larger range in an address space, reducing the number of TLB misses. 
A smaller number of misses reduces the address translation overhead.
%This significantly reduces translation overheads by increasing TLB coverage.
% 
}

Big-memory workloads are well known to witness significant slowdowns due to virtual to physical address translation (e.g., up to 20-50\%).
Modern processors support large pages to help reduce this overhead~\cite{panwar:asplos:2018, panwar:asplos:2019, kwon:osdi:2016}.
A large page TLB (Translation Lookaside Buffer) entry maps a larger contiguous virtual address region to contiguous physical address region (e.g., 2MB or 1GB compared to default 4KB).
Consequently, the use of a large page increases TLB coverage and can reduce the number of TLB misses that is the primary source of translation overheads.

%\newtext{
However, the hardware support \textit{alone} is not useful. 
%The OSes and hypervisors (henceforth called \textit{system software}) that are responsible for creating the address mappings need to enable large pages for applications.
The OSes and hypervisors (henceforth called \textit{system software}) that create the address mappings need to enable large pages for applications.
%Further, the ease of using large pages as enabled by the system software determines how prevalent the use of large pages would be in practice.
Further, the ease of use determines the prevalence of large page's deployment in practice.
%For example, if significant modifications to applications are necessary to use a given large page size, the usage of the hardware support for that large page size is likely to be limited.
If application modifications are necessary for allocating a given large page size, then that page size may not be widely deployed and, consequently, render the corresponding hardware resources underutilized. 
%Consequently, the hardware resources for that  page size would remain underutilized.
%Applications cannot leverage the hardware support for large page, unless system software e
%For example, the page sizes are fixed by the hardware, but a new page size is useless without system software enabling it to the software ecosystem.
%}

The x86-64 processors have supported two large page sizes -- 2MB and 1GB, for over a decade. Intel Sandybridge architecture launched in 2010 supported 1GB pages and had a four-entry L1 TLB dedicated for 1GB pages in each core~\cite{wikichip:sandybridge}.
The current generation of Intel Coffee Lake processors additionally have a 16-entry L2-TLB for 1GB pages~\cite{wikichip:coffeelake}.
Upcoming Intel Ice Lake processors is presumed to have 1024-entry L2 TLB for 1GB pages~\cite{anandtech:icelake:2019}.
In short, processor vendors continues to enhance support for 1GB pages.

\ignore{
It is thus imperative to look back at the added hardware support to analyze its usefulness to applications and possibly enhance the software support to drive up hardware's usefulness.}

Unfortunately, software enablement of large pages has focused primarily on 2MB large pages. 
Linux's Transparent Huge Page (\thp{}) enables dynamic allocation of large pages without user intervention and thus, is key to the widespread use of large pages.
But it is limited only to 2MB pages.    
%Linux currently does not allow dynamic allocation of 1GB pages.
Previous research works on improving large page support similarly ignore 1GB pages~\cite{kwon:osdi:2016, panwar:asplos:2018, panwar:asplos:2019}.

With the continued growth in the memory footprint of applications, use of 1GB pages is likely to become a necessity for good performance.
The advent of denser non-volatile memory (NVM) technologies promises to significantly increase the physical memory size~\cite{optane:usecases:2019,intel:3dxpoint}. 
The ability to efficiently address a large amount of memory is essential to harness the benefits of NVM.
While the hardware support for 1GB pages has sped ahead, its software enablement has fallen behind.

However, it is important for architects to find hard evidence of practical usefulness of 1GB pages, over and above 2MB pages, to continue enhancing its support or else consider dropping it in future products. 
Therefore, we first set out to quantify the usefulness of 1GB pages to various applications, with and without virtualization. 
%Guided by this analysis, we then enhance Linux to deploy all three page sizes available in x86 processors for memory-intensive applications without user intervention. 
%We start by empirically studying a wide range of applications with large memory footprints to understand which applications may benefit from 1GB pages, both with and without virtualization. , both 
We find that while most memory-intensive applications benefit from 2MB pages over 4KB, a subset of them speeds up further with 1GB pages.
%Under virtualization, the usefulness of 1GB pages slightly increases. 
%However, using only the largest available page size (here, 1GB) is not always optimal. 
Even for applications in that subset, use of the \textit{other} large page size(s) (here, 2MB) \textit{alongside} the largest page size (here, 1GB) is important.
Mapping a virtual address range with a large page size requires the address range to be at least as long as that page size and be aligned at that page size boundary. 
Consequently, larger the page size, lesser is the number of virtual address ranges that are \textit{mappable} by that page size.  
%allocating 1GB pages requires contiguously allocated virtual address ranges which are 1GB or longer and have starting addresses that are 1GB-aligned.
%2MB pages require only 2MB or longer regions that are 2MB-aligned.

We empirically find that often a significant portion of an application's address space is \textit{not} mappable with 1GB pages, but mappable with 2MB.
Importantly, such address ranges often witness relatively frequent TLB misses.
Thus, if \textit{only} the largest page size is used, then those address ranges would have to be mapped with the smallest (4KB) pages. Consequently, the number of TLB misses would increase significantly.

A larger page size also needs equally longer contiguous physical memory chunk.
The larger the page size, shorter is the supply for necessary physical memory chunks.
Thus, it may not be possible to map a virtual address range with the largest page size even if it is mappable by that page size.
%1GB contiguous physical memory required to map a 1GB page may be unavailable, but 2MB contiguous regions are available. 
%Thus if 1GB contiguous physical memory is available, one can fall back to 2MB pages if allthree page sizes }are used. 
%Driven by this analysis, we enable dynamic allocation of \textit{all available page sizes} on x86-64 processors. 

%Further, kernel itself is one of the significant users of 1GB pages.
%Linux, for example, uses 1GB pages to map the entire physical memory of the system into its kernel's virtual address space during boot (called \textit{direct-mapping}) to minimize TLB misses during the kernel execution.
%However, our study suggests that the use of 1 GB pages may not be always beneficial to Linux's performance, it can even be detrimental.

%The use of 1GB pages is particularly hamstrung by the inability allocate them dynamically without user intervention.
%Linux's \thp{} supports \textit{only} 2MB pages.
%We thus, designed and implemented \trident{} in Linux to dynamically allocate all three available page sizes. 
%Like \thp{}, \trident{} is transparent to applications. 
%We thus implemented an application-transparent dynamic allocation of 1GB pages in Linux to bridge this gap.
%To the best of our knowledge, \trident{} is the first framework to dynamically allocate three different page sizes in x86-64 based systems. 

%we created first OS framework that could dynamically map an application's virtual address space using three different page sizes -- 4KB, 2MB and 1GB, as deemed suitable.   

Driven by the above analysis, we built \trident{} in Linux to dynamically allocate \textit{all available page sizes} in x86-64 systems. 
A key challenge in dynamically allocating 1GB pages is the lack of enough number of 1GB \textit{contiguous} physical memory chunks.
As the free physical memory gets naturally fragmented over time, finding 1GB chunks become far more difficult than 2MB chunks.
%In particular, we find many practical challenges to dynamically allocate 1GB pages.  
%In particular, we observe that the use of 1GB pages is hamstrung by the inability to allocate them without user intervention.
%While \thp{} has long been deployed in Linux for 2MB pages, extending it to utilize all page sizes is anything but straightforward.
%For example, when the physical memory is fragmented,
%(i.e., free memory is scattered in smaller holes),
%finding \textit{contiguous} free physical memory of 1GB or longer is significantly harder than 2MB.
%Allocating 1GB page requires 1GB of , finding which is significantly harder than finding a 2MB region.
%This is especially true when the physical memory is fragmented (i.e., free memory is scattered in smaller holes).
%Fragmentation happens naturally over time as applications allocate/de-allocate memory. 
Thus, the dynamic allocation of large pages needs to periodically compact physical memory to make contiguous free memory available.
%Unfortunately, uninformed sequential scanning to move contents of the mapped (occupied) base (4KB) pages as done by \thp{} is inefficient for 1GB pages.
%Overhead of compaction for 1GB pages could negate much benefits of saving TLB misses by 1GB pages.
However, compaction for 1GB memory chunk requires significantly more work than 2MB.
%However, creating even a \textit{single} contiguous 1GB region requires significantly more work
%moving a much larger number of bytes and scanning of larger portions of memory, compared to 2MB.
Moreover, a compaction attempt fails if it encounters even a \textit{single} page frame (4KB) with unmovable contents, e.g., kernel's objects like \textsf{inodes}, in a 1GB region. 
%Moreover, if a 1GB region contains even a single page frame with unmovable contents, e.g., kernel's objects like \texttt{inodes}, then all work performed to compact that region till encountering the unmovable page is wasted~\cite{panwar:asplos:2018}.
In short, compaction for 1GB chunks needs a \textit{new} approach. 

\ignore{
This shows up as significant CPU utilization due to compaction activity. 
Consequently, the applications threads gets a reduced fraction of CPU cycles with more CPU cycles consumed by the kernel.
}

\trident{} introduces a novel \smartcompaction{} technique.
We observe that the current compaction approach of \textit{sequential scanning} and moving contents of occupied page frames
%(as in \thp{} for 2MB pages)
is not scalable to 1GB.
This approach incurs an unnecessarily large amount of data movement. 
%This approach moves the contents of an unnecessarily large number of pages.
The \smartcompaction{} tracks the number of occupied bytes  (i.e., the number of mapped page frames) within each 1GB physical memory chunk.
Instead of scanning, \smartcompaction{} then frees a region with the \textit{least} number of occupied bytes, which significantly reduces data movement.
%Consequently, overhead of compaction reduces.
It also tracks unmovable contents within a 1GB region to further avoid unnecessary data movement.
%It also decreases the probability of encountering a page with unmovable content that would have otherwise resulted in failure to compact. 

Even with \smartcompaction{}, 1GB memory chunks are not always available when needed.
About a third of the attempts to allocate 1GB page fails due to the unavailability of contiguous physical memory.
%It is thus important to use \textit{all three page sizes} -- not only 1GB pages.
%But a failed attempt to compact for a 1GB region could create many 2MB regions. 
Unsurprisingly, 2MB chunks are more easily available. 
\trident{} thus maps address ranges with 2MB pages if it fails to map with 1GB pages. 
Later, these 2MB page mappings are \textit{promoted} to 1GB pages, when suitable.
%It then periodically attempts to later promote these 2MB mappings to 1GB when and where possible.

\ignore{\trident{}'s implementation had to address several other shortcomings in Linux. 
%For example, Linux's physical memory allocator
%management (e.g., buddy allocator) is designed to 
%maintains only up to 4MB of physical memory chunks~\cite{buddy}.
%\trident{} extends Linux to manage up to 1GB memory chunks for dynamically allocating them. 
For example, allocating a 1GB page during page fault is $500\times$ slower than a 2MB page due to overheads of zero-filling memory.
\trident{} employs asynchronous zero-filling to speed up 1GB allocations
for reducing tail latency.}

%Linux's NUMA subsystem also needed to be made aware of 1GB pages. 

We then propose \tridentpv{}, an \textit{optional} extension to \trident{} under virtualization for copy-less 1GB page promotion and compaction in the guest OS.
The guest often copies contents of guest physical pages to create contiguity in guest physical address space for page promotion and compaction.
We observe that coping can be \textit{mimicked} by exchanging the mapping between the guest physical address (\textsf{gPA}) and the host physical address (\textsf{hPA}) of the source and destination.
This copy-less technique makes the promotion of 2MB pages to a 1GB page significantly faster than the traditional copy-based approach.
However, the guest and the hypervisor need to coordinate to alter the desired \textsf{gPA} to \textsf{hPA} mappings via a hypercall.

We find that on a bare-metal system, \trident{} speeds up eight memory-intensive applications by $18\%$, over Linux's \thp{}, on average. \tridentpv{} can improve upon \trident{}'s own performance under virtualization, by up to $10\%$.  

In short, this paper shows that the use of 1GB pages has been hamstrung due to inadequate software enablement even after a decade of hardware support in commercial processors. 
Therefore, architects must prioritize enhancing the software support before investing further into hardware support for 1GB pages. Our specific contributions are as follows:

%Our contributions are as follows:
\vs{-1mm}
\begin{itemize}
    \item We evaluate the usefulness of 1GB pages across various applications, both without and with virtualization.
    
    %\item \newtext{We demonstrate that use of 1GB pages is hamstrung by lack of adequate software enablement and before enhancing hardware support further, architects should enhance its software support first.} 
    %, and for the Linux kernel itself. 
    \item We empirically demonstrate why it is important to deploy \textit{all} large page sizes, not only the largest one. 
    %\item We find that application-transparent allocation of 1GB pages could increase utilization hardware resources invested for 1GB pages.
    \item We created \trident{} in Linux to dynamically allocate all page sizes available in x86-64 processors to significantly speedup applications with large memory footprint.
    
    \item We then propose an optional extension to \trident{} called \tridentpv{} that employs paravirtualization to enable copy-less 1GB page promotion and compaction in the guest.
    %We will open source \trident{} upon publication of this work and expect it to help applications better leverage currently underutilized hardware support 1GB pages. 
\end{itemize}

%\newtext{Preferred location for a small paragraph on ``lessons for architects".}

\section{Background}
\label{sec:background}

%We discuss the importance of large pages and summarize different mechanisms available to allocate large pages.

%\subsection{Hardware support for large pages}
%\label{subsec:hardware_large_pages}
\noindent{\bf Hardware support for large pages:}
Applications with large memory footprints often spend considerable time in address translation (e.g., up to 20-50\%)~\cite{basu:isca:2013, panwar:asplos:2019}.
Address translation overhead is primarily the overhead of performing page table walks on TLB misses.
Hits in the TLB are fast, but a page table walk on a miss may require up to four memory accesses to lookup the hierarchical in-memory page table in x86-64 processors. 
%The hardware support for large pages is designed to reduce this overhead and is prevalent across most processor architectures. 
Large pages can help reduce translation overhead in two ways:
\circled{1} It reduces the frequency of TLB misses by increasing the TLB coverage since a single entry for a large page maps a larger address range. 
\circled{2} It quickens individual page table walks by reducing the number of levels in the page table that need to be looked up.
For example, a walk for a 1GB page requires up to 2 memory accesses, compared to 3 for a 2MB page and 4 for a 4KB page, in x86-64 processors. %In addition, large pages minimize memory consumed by translation structures (e.g., page tables) leading to more effective cache utilization.

\ignore{Further, recent work has shown that page walk overheads are exacerbated by NUMA architectures due to higher latency of remote memory accesses~\cite{mitosis}. Current systems employ various techniques to optimize performance on NUMA systems. For example, Linux prefers node-local memory allocation to minimize remote memory accesses and uses additional
kernel services (e.g., AutoNUMA ~\cite{autonuma}) for migrating memory pages to optimize data placement at runtime. However, current systems do not support page table migration and hence \textit{remote} page walks can significantly deteriorate performance.}

%\subsection{Large pages under virtualization}
%\label{subsec:large_pages_virt}
\noindent
{\bf Large pages under virtualization:}
Address translation involves two layers under virtualization through nested page tables.
First, a guest virtual address (\textsf{gVA}) is mapped to a guest physical address (\textsf{gPA}) through the guest page tables (\textsf{gPT}) managed by the guest OS running on a virtual machine.
\textsf{gPA} is then translated to host physical address (\textsf{hPA}) through host page tables (\textsf{hPT}) maintained by the hypervisor. 
Two layers of indirection increase the number of memory accesses required for a page walk. 
For example, with four-level page tables, a TLB miss requires up to 24 memory accesses for 4KB-pages.
Use of 2MB and 1GB pages at both layers reduce the number of accesses to 15 and 8, respectively.
\noindent
{\bf OS support for large page allocation:}
OSes typically provide three mechanisms to allocate
large pages. 
In the pre-allocation based mechanism, users are required to reserve physical memory for large pages and a helper library (e.g., \textsf{libHugetlbfs}) maps specific segment(s) of an application's memory with large pages from the reserved memory. 
Unfortunately, this \textit{static} approach constrains the usability of large pages.
The second approach needs explicit system calls to map a virtual address ranges with large pages.
This requires application modification (e.g., \textsf{madvise} syscall or extra flags in \textsf{mmap}). 
%It may not be always easy for developers to determine suitable page sizes for a given address range.
In the third approach, the OS allocates large pages \textit{without} the user or programmer involvement.  
Linux's Transparent Huge Pages (\thp{}) is an example of this approach.
%as the user needs  apriori decide the number of large pages and program segment to use them for.
%constraints dynamic memory management as freeing/allocating memory requires
%user intervention.

%With \thp{}, the OS can transparently allocate large pages to applications without explicit
%application hints or pre-allocation of memory. 
%\thp{} also enabled
%dynamic memory management as the OS can reclaim memory from large pages when required
%(e.g., under low memory). 
%In this paper, we primarily focus on transparent
%huge pages as this are the most widely accepted mode of large page support.

Internally, \thp{} employs two mechanisms.
On a page fault, it checks if the faulting address falls within a virtual address range that is at least as big as and aligned with the large page size. 
If yes, and a free contiguous physical memory chunk is available, \thp{} maps the address with a large page.
For address regions that were not immediately mappable with large pages during page faults, \thp{} employs a background thread (\textsf{khugepaged}) to locate virtual address ranges mapped with 4KB pages and {\it promote} (remap)
them to large pages, when possible.
To ensure enough supply of contiguous physical memory, \thp{} also \textit{compacts} physical memory.
Compaction moves contents of occupied pages to one end of the physical memory for creating contiguous free memory regions on the other end. Unfortunately, \thp{} currently supports \textit{only} 2MB large pages.
Hosted hypervisors, such as \textsf{\small KVM} uses \thp{} for allocating 2MB large pages in the hypervisor (host).
%VMware ESX server also employs technique similar to \thp{} for large page allocation in the hypervisor~\cite{esxi:thp}.

\ignore {
\subsection{Large pages under virtualization}
\label{subsec:large_pages_virt}
Under virtualization, there are two levels of address translations.
The (guest) applications running within a virtual machine accesses memory using guest virtual address (gVA).
To complete a memory access (e.g., load/store) issued by an guest application the gVA has to be first translated to guest physical address (gPA). 
The page table storing the mapping between gVAs and gPAs is called guest page table (gPT) and is maintained by the guest OS running on a virtual machine.
The gPA of a memory access is further needs to be translated to system or real physical address (sPA).
The mapping between gPA and sPA is stored in the system page table (sPT) and is maintained by the hypervisor. 
Due to this two-level translation the overheads of address translation is often higher under virtualization. 
Furthermore, it is possible to use different page sizes for translation in the two levels and thus a total of nine possible page size combinations are possible.
}

\section{Methodology}
\label{sec:methodology}

\begin{table}[]
\caption{Specification of the experimental system}
\center
\scalebox{0.80}{
\begin{tabular}{|l|l|}
\hline
\textbf{Processor} & \begin{tabular}[c]{@{}l@{}}Intel Xeon Gold 6140  @2.3GHz\\ Skylake Family with 2 Sockets\end{tabular}                             \\ \hline
\textbf{Number of cores}     & 18 cores (36 threads) per socket \\ \hline
\textbf{L1-iTLB}   & \begin{tabular}[c]{@{}l@{}}4KB pages, 8-way, 128 entries\\ 2MB pages, fully associative, 8 entries\end{tabular}                    \\ \hline
\textbf{L1-dTLB}   & \begin{tabular}[c]{@{}l@{}}4KB pages, 4-way, 64 entries\\ 2MB pages, 4-way, 32 entries\\ 1GB pages, fully associative, 4 entries\end{tabular} \\ \hline
\textbf{L2 TLB}    & \begin{tabular}[c]{@{}l@{}} 4KB/2MB pages, 12-way, 1536 entries \\ 1GB pages, 4-way, 16 entries \end{tabular} \\ \hline
\textbf{Cache}     & 32K L1-d, 32K L1-i, 1MB L2, 24MB L3 \\ \hline
\textbf{Main Memory}    & 384GB (192GB per socket)  \\ \hline
\textbf{OS / Hypervisor}  & Ubuntu with Linux kernel version 4.17.3 / KVM \\ \hline
\end{tabular}
}
\label{table:systemspecs}
\end{table}

\ignore{
\subsection{Experimental setup}
\label{subsec:setup}
Our experimental platform is a two-socket Intel Skylake server system
with Intel Xeon Gold 6140 processors running with Ubuntu16.04. Each
socket contains 36 threads (with hyperthreading enabled) and 192GB memory.
The L1 TLB contains 64 and 8 entries for 4KB and 2MB pages respectively
while the L2 TLB contains 1024 entries for both 4KB and
2MB pages. The size of L1, L2 and shared L3 cache is 768KB, 3MB
and 30MB resp.

\subsection{Workloads}
\label{subsec:workloads}
Our evaluation involves a diverse set of workloads ranging from HPC, graph algorithms,
in-memory databases, genomics and machine learning
\cite{redis,xsbench,nas,gapbs,parsec,biobench,cpu2006,graph500}.
HawkEye is implemented in Linux kernel v4.3.
}
\autoref{table:systemspecs} details the configuration of our experimental platform. 
We evaluate 12 workloads with multi-GB memory footprint (see \autoref{table:benchmarkspecs}) across machine-learning, graph algorithms, key-value stores, HPC, and microbenchmarks ({\sf \small GUPS} and {\sf Btree}).
We use Linux's perf tool~\cite{perf} to collect microarchitectural events related to virtual memory overheads. 
Specifically, we monitor events like the number of cycles spent on page walks via counters \textsf{\small DTLB\_LOAD\_MISSES.WALK\_DURATION} and  \textsf{\small DTLB\_STORE\_MISSES.WALK\_DURATION}.
%and total execution cycles using \textsf{\small CPU\_CLK\_UNHALTED} counter. 

To study performance under different states of the system, we used a tool to fragment the physical memory (more in Section~\ref{subsec:design}).
Physical memory is fragmented by reading a large file at random offsets to populate OS's file cache.
This renders physical memory sprinkled with recently used page frames containing file cache contents and thus, fragmented.
%The tool is similar in spirit to one used in a recent publication~\cite{panwar:asplos:2019}.

%We performed the experiments under virtualization using Linux's Kernel Virtual Machine or \texttt{KVM}~\cite{kvm}. 
%The guest OS is the same Linux version (See Table~\ref{table:systemspecs}) as used in the host. 

\begin{table}[t!]
\caption{Specifications of the benchmarks}
\center
\scalebox{0.70}{
\begin{tabular}{|l|r|r|l|}
\hline
\textbf{Name} & \textbf{Threads} & \textbf{Memory} & \textbf{Description} \\ \hline
XSBench & 36 & 117GB & \begin{tabular}[c]{@{}l@{}}Monte Carlo particle transport algorithm \\ for nuclear reactors~\cite{xsbench}\end{tabular}                                     \\ \hline
SVM & 36 & 67.9GB & Support Vector Machine, kdd2012  dataset ~\cite{svm-dataset} \\ \hline
Graph500 & 36 & 63.5GB & \begin{tabular}[c]{@{}l@{}}Breadth-first-search and single-source-shortest-\\path over undirected graphs~\cite{graph500}\end{tabular} \\ \hline
CC/BC/PR & 36 & 72GB & Graph algorithms from GAPBS~\cite{gapbs} \\ \hline
CG.D & 36 & 50GB & \begin{tabular}[c]{@{}l@{}}Congruent Gradient algorithm from NAS \\ Parallel Benchmarks~\cite{nas}\end{tabular}  \\ \hline
Btree & 1 & 10.5GB & Random lookups in a B+tree  \\ \hline
GUPS & 1 & 32GB & Irregular, memory-intensive microbenchmark~\cite{gups} \\ \hline
Redis & 1 & 43.6GB & An in-memory key-value store~\cite{redisbook} \\ \hline
Memcached & 36 & 79GB & An in-memory key-value caching store~\cite{memcached} \\ \hline
Canneal & 1 & 32GB & Simulated cache-aware annealing from PARSEC~\cite{parsec}\\ \hline
\end{tabular}
}
\label{table:benchmarkspecs}
\end{table}

\ignore{
a simple script to create fragmentation in the physical memory for conducting experiments under fragmentation.
This script reads a very large file in a random fashion.
This triggers the OS to allocate memory for its file cache.
The physical page frames used for file cache gets scattered over the physical memory to create significant fragmentation.
}

\ignore{
As described in the Section~\ref{sec:methodology}, we implemented \trident{} in Linux and evaluated it on a server with Intel processors. We used Linux's perf tool~\cite{perf} to collect hardware counter data. 
More specifically, we use \todo{counters....}.

We used a simple script to create fragmentation in the physical memory for conducting experiments under fragmentation.
This script reads a very large file in a random fashion.
This triggers the OS to allocate memory for its file cache.
The physical page frames used for file cache gets scattered over the physical memory to create significant fragmentation.
}
\vspace{-0.5em}
\section{How useful are 1GB large pages?}
\label{sec:usefullness}
%While we observe relatively sparse usage of 1GB pages in the software, all processors pay the cost of supporting 1GB pages. 

Hardware support for 1GB pages is not free, and the software running on x86-64 processors pays the price, irrespective of its use of 1GB pages.
For example, modern Intel processors have 4-entry L1 TLB and 16-entry L2 TLB dedicated to 1GB pages. 
Those four L1 entries for 1GB pages are accessed on \textit{every} load and store since the page size is not known during TLB lookup. 
%Effectively, 1GB L1 TLBs increase the associativity of L1 TLB lookup from eight to twelve (a $50\%$ increase). 
Due to frequent accesses, L1 TLBs can contribute to a thermal hotspot in processors~\cite{puttaswamy:gvlsi:2006} and can account for $6\%$ of a processor's total power~\cite{sodani:micro:2011}.  
The presence of dedicated TLBs for 1GB pages adds to the cost.
The continued increase in the number of TLB entries for 1GB pages would worsen it.

It is, thus, natural to wonder if applications can benefit from 1GB pages.
We analyze various applications under different execution scenarios to understand usefulness of 1GB pages.

\begin{figure}[t]
    \centering
    \includegraphics[width=\columnwidth]{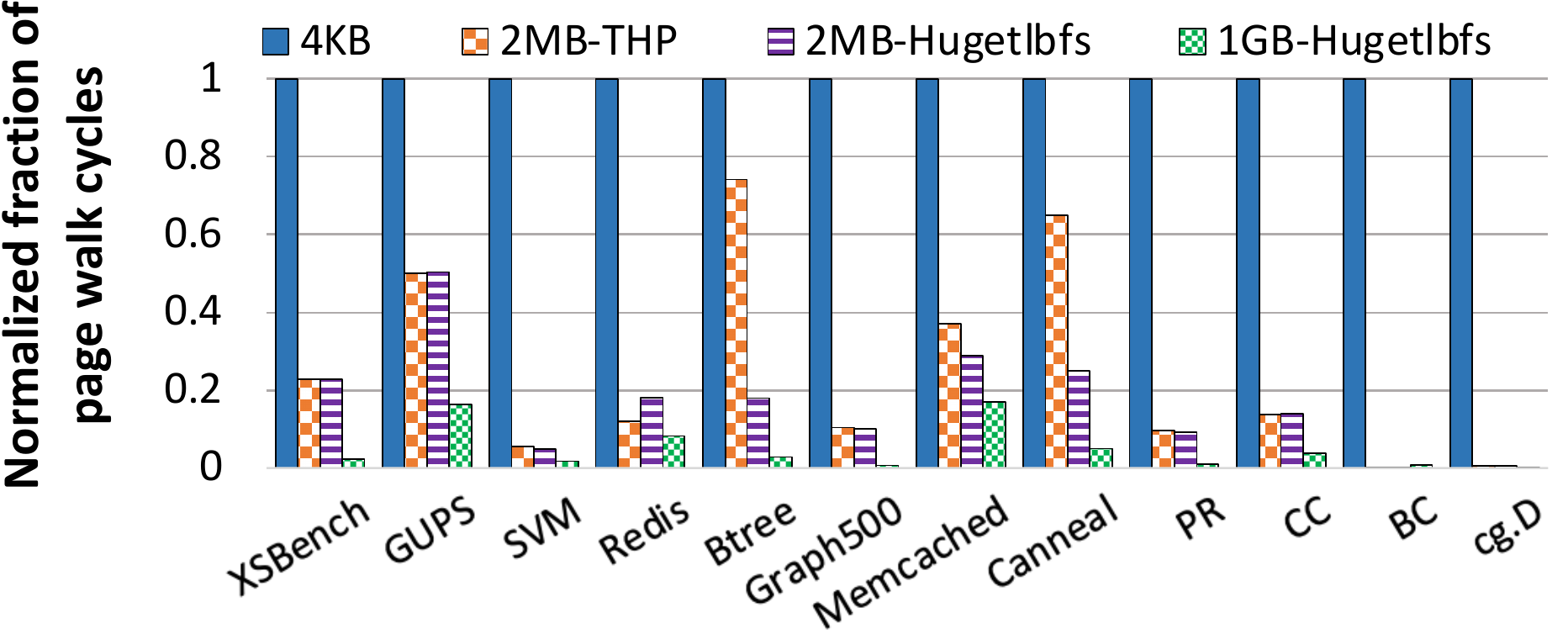} 
    \vs{-1.5em}
    \caption{Fraction of page walk cycles in native execution.}
    \vs{-1.5em}
    \label{fig:unvirt_num_walk}
\end{figure}

\begin{figure}[t]
    \centering
    \includegraphics[width=\columnwidth]{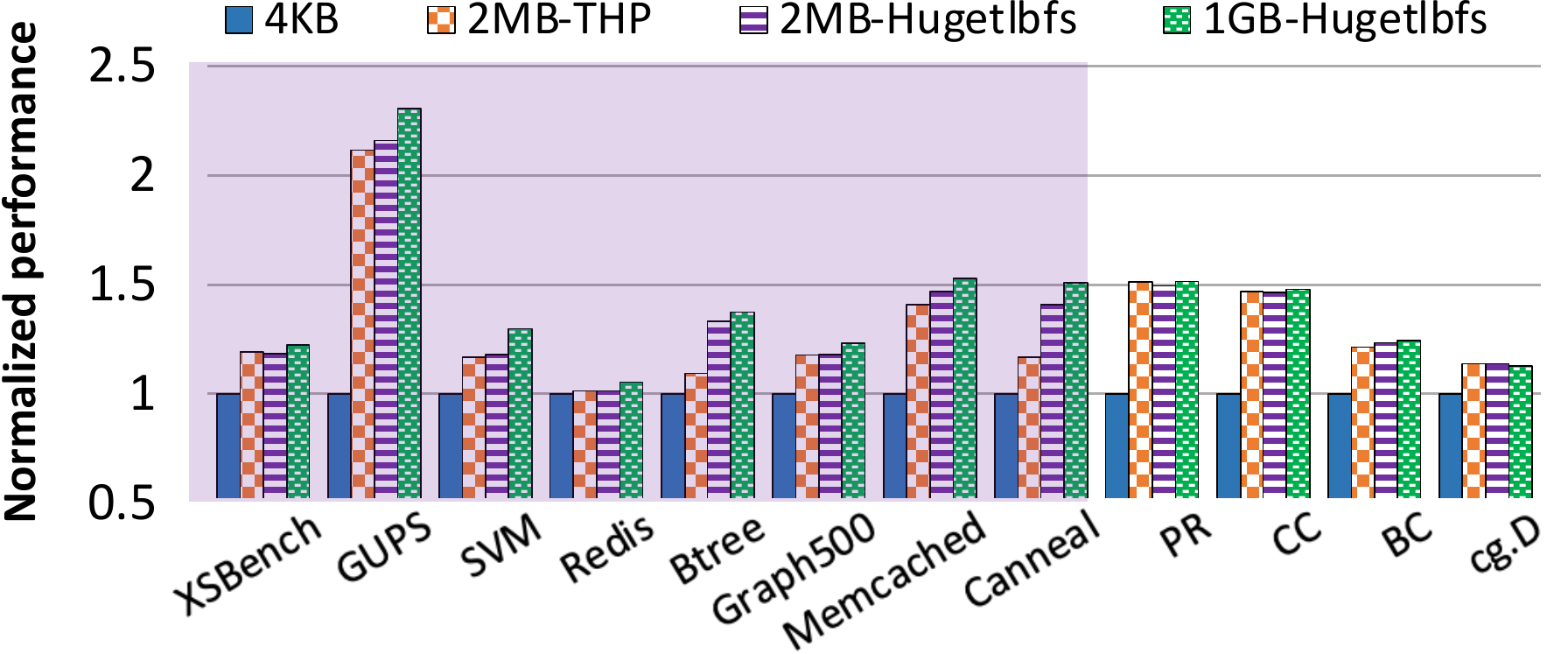}
    \caption{Performance under native execution. Applications in shade benefit from 1GB pages.}
    \label{fig:unvirt_perf}
\end{figure}

\subsection{1GB pages in native execution}
\label{subsec:analysis_univirt}
\autoref{fig:unvirt_num_walk} shows the (normalized) fraction of execution cycles spent on page walks for each application while using different page sizes.
The four bars for each application represent walk cycles with \circled{1} 4KB pages, \circled{2} dynamically allocated 2MB pages via \thp{}, \circled{3} statically pre-allocated 2MB pages via \textsf{libHugetlbfs}, and, \circled{4} statically pre-allocated 1GB pages via \textsf{libHugetlbfs}. 
The fourth bar {\it approximates} the performance achievable if the 1GB pages are deployed but not 2MB.   
Note that the application-transparent dynamic allocation for 1GB pages (i.e., \thp{} like) is not supported in Linux today.

We note that Linux's \thp{} often performs as good as 2MB-\textsf{libHugetlbfs}.
For \textsf{Redis}, \thp{} reduces more walk cycles than \textsf{libHugetlbfs}.
We find that this is because \textsf{Redis}'s stack memory is TLB sensitive, which cannot be mapped using \textsf{libHugetlbfs}.
Importantly, \thp{} does not require pre-allocation of physical memory, nor does it need users to statically decide which program segment(s) to be mapped with large pages.

\begin{figure}[t]
    \centering
    \includegraphics[width=\columnwidth]{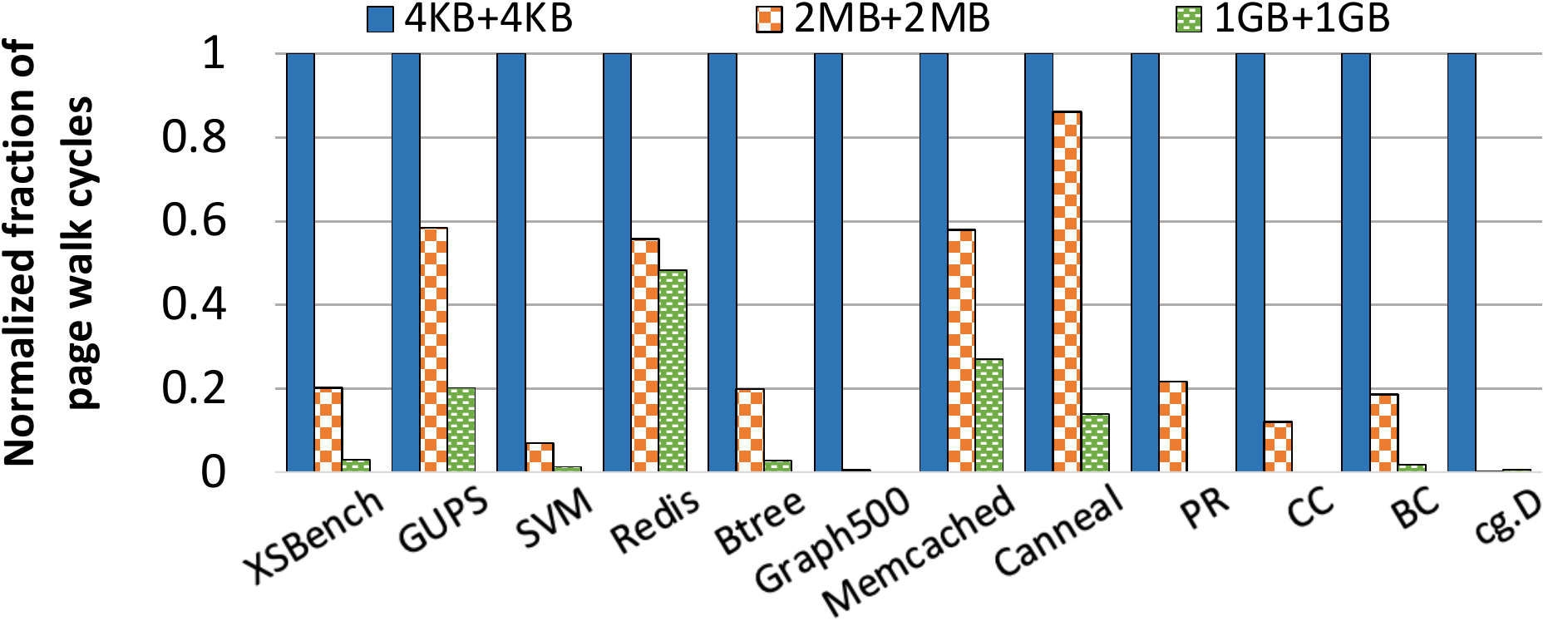}
    \vs{-1.5em}
    \caption{Fraction of page walk cycles under virtualization.}
    \vs{-1.5em}
    \label{fig:virt_page_walks}
\end{figure}

\begin{figure}[t]
    \centering
    \includegraphics[width=\columnwidth]{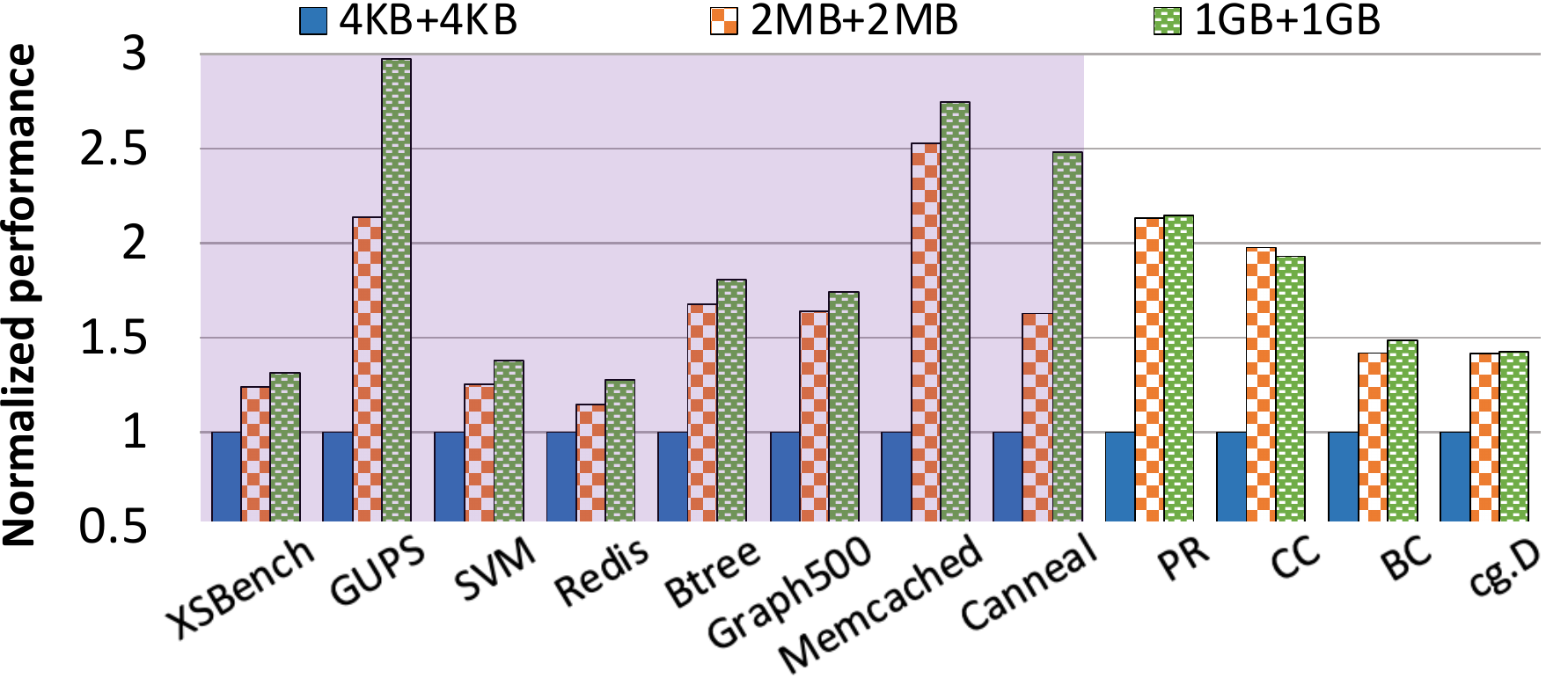}
     \caption{Normalized performance under virtualization.}
    \label{fig:virt_perf}
\end{figure}

Reductions in walk cycles do not necessarily translate to proportional performance improvement on out-of-order CPUs.
It depends upon what portions of those cycles are in the critical path of the execution. \autoref{fig:unvirt_perf} shows the normalized performance.
For all workloads, except \textsf{Redis}, the performance is calculated as the inverse of the execution time.
For \textsf{Redis}, the performance is the throughput as reported by the benchmark. 
%Each application has four bars as in the previous figure.
%The height of each bar is normalized to that with 4KB pages (higher is better).

We observe non-negligible performance improvement (at least $3\%$) for eight applications (shaded left part of the figure) from 1GB pages (over 2MB).
For example, \textsf{Canneal} speeds up by $30\%$ over \thp{}.
These eight applications' performance improves by $12.5\%$, on average, when 1GB pages are used via \textsf{libHugetlbfs}, relative to \thp{} using 2MB pages.  
Rest of the  applications witness benefits of using 2MB pages (over 4KB), but barely gain any further with 1GB pages. 
This is not surprising; the walk cycles were already low with 2MB pages, and an out-of-order CPU could hide the remaining cycles.  
\ignore{
Moreover, a 2MB page walk latency is typically lower than that for a 4KB page walk as three, instead of four levels of the page table are required to be looked up. 
Thus, eliminating individual 2MB page walks with 1GB pages saves a relatively smaller number of cycles in the critical path in comparison to saving 4KB page walks by 2MB pages. 
These smaller number of cycles could already be hidden by large out-of-order cores.  }

For the rest of the paper, we will thus focus on the first eight (shaded) applications. %(\textsf{XSBench, GUPS, SVM, Redis, Btree, Graph500, Memcached, Canneal}).
We also observe that \thp{} is able to perform within $0.5\%$ of that of \textsf{libHugetlbfs} using 2MB pages without needing memory pre-allocation or user guidance.
This emphasizes the importance of \thp{} in the wide deployment of 2MB pages; something yet to be realized for 1GB pages.

\ignore{In summary, we observe that performance of five out of nine applications with large memory footprints improve appreciably with 1GB pages (in the shaded left half of the figure).}

\ignore{
\noindent\fbox{
\begin{minipage}{25em}
Observations:
\begin{itemize}
\setlength{\itemindent}{-2em}
    \item \noindent Only a handful of applications experience significant performance uplift relative to 2MB pages due to use of 1GB pages, although many applications . 
    \item \noindent Application-transparent allocation of 2MB pages (a.k.a., THP) brings almost all benefit of 2MB pages without requiring user or programmer intervention. Unfortunately, such application-transparent allocation for 1GB pages does not currently exists. 
\end{itemize}
\end{minipage}
}}

\subsection{1GB pages under virtualized execution}
\label{subsec:analysis_virt}
Two levels of translation under virtualization can increase overheads.
Each level may use a different page size.
Thus, a total of nine combinations of page sizes are plausible.
While we experimented with all, we discuss only 4KB-4KB, 2MB-2MB, and 1GB-1GB combinations.
The first term denotes the page size used for guest, and the second term denotes that in the host. 
We chose these configurations as they demonstrate the best performance achievable with a given page size.

\autoref{fig:virt_page_walks} shows the normalized fraction of page walk cycles under three different page size combinations.
%There are three bars for each application -- one for configuration. 
We notice significant reductions in walk cycles with 2MB and 1GB pages.
For example, the fraction of walk cycles reduced by $80\%$ for \textsf{XSBench}. 
Even a couple of 1GB page agnostic applications e.g., \textsf{PR} and \textsf{CC} experience a large reduction in walk cycles.

\autoref{fig:virt_perf} shows the performance under virtualization. 
We observe that 1GB pages provides a bit more benefit here. 
The eight 1GB page sensitive applications speed up by $17.6\%$ over 2MB pages, on average.
The application \textsf{BC}, which did not benefit from 1GB pages under native execution, becomes slightly sensitive to 1GB pages. 

\begin{figure}[t]
\centering
%\begin{tabular}{@{}c@{}c}
\begin{tabular}{@{}c@{}c}
    \includegraphics[scale=.24]{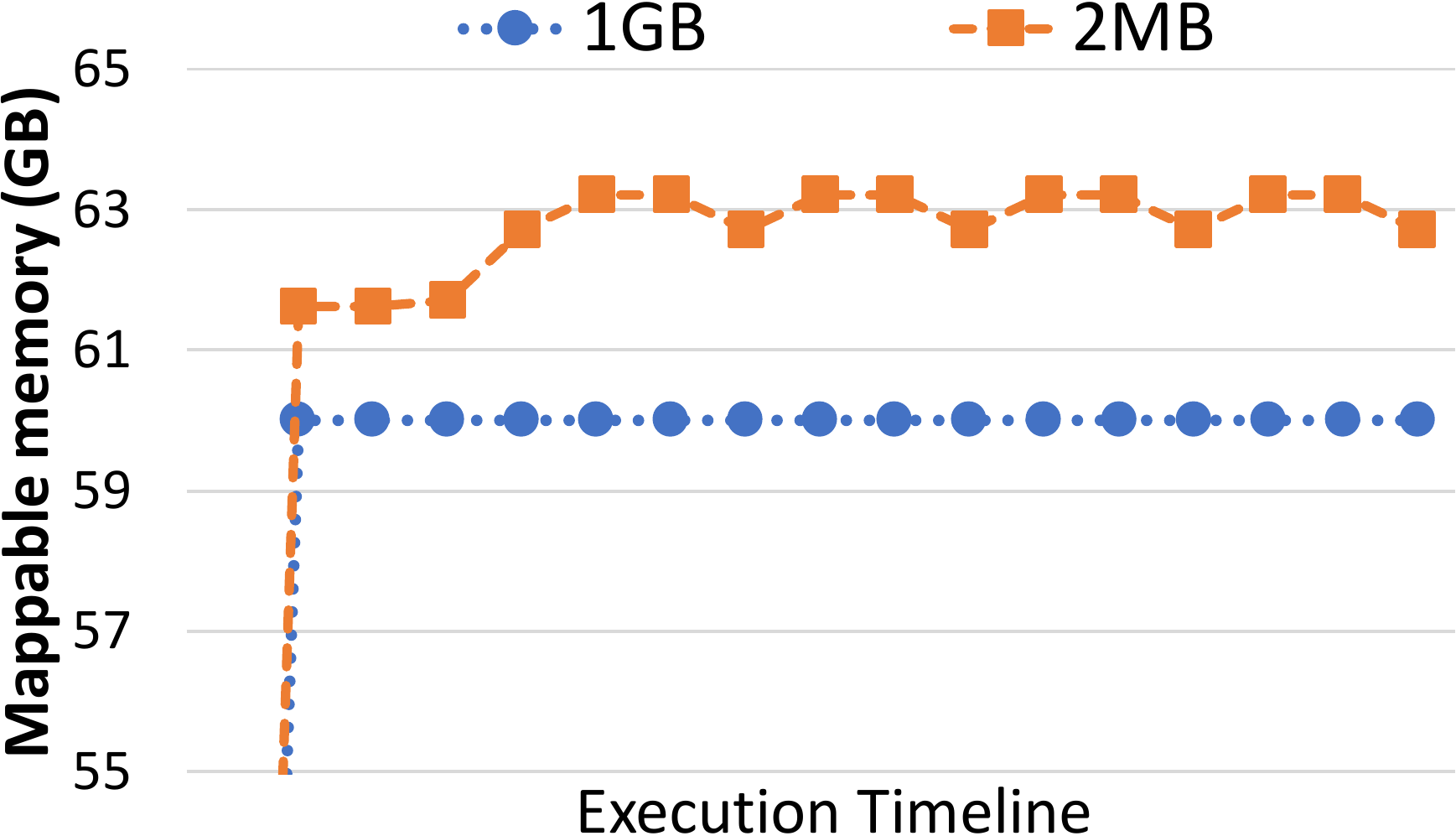} &  \includegraphics[scale=.24]{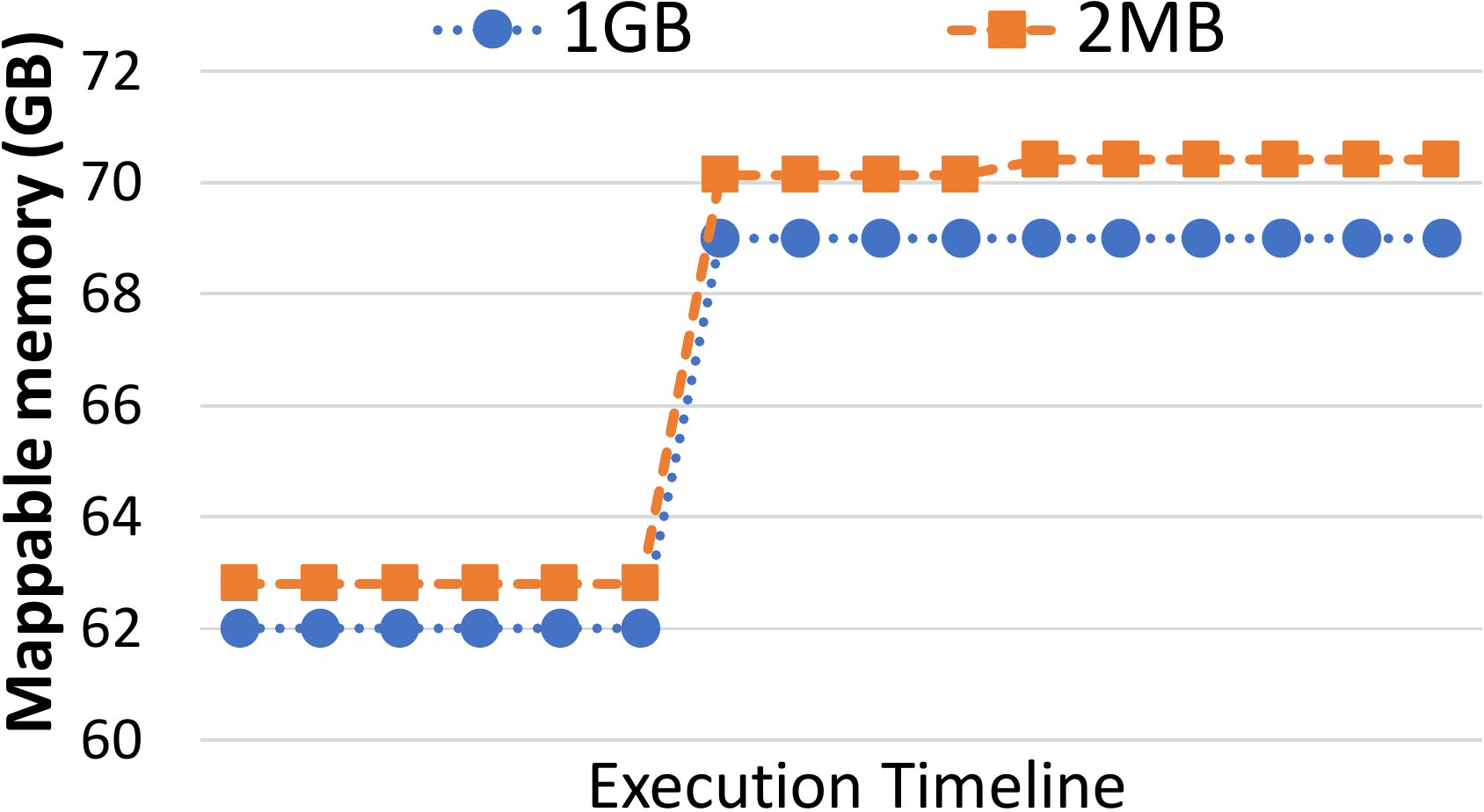} \\
    (a) Graph500 & (b) SVM \\
\end{tabular}
    \caption{Total memory mappable with different page sizes.}
    \vs{-1.5em}
    \label{fig:mappable}
\end{figure}

\begin{figure}[t]
\centering
%\begin{tabular}{@{}c@{}c}
\begin{tabular}{@{}c@{}c}
    \includegraphics[width=0.5\columnwidth]{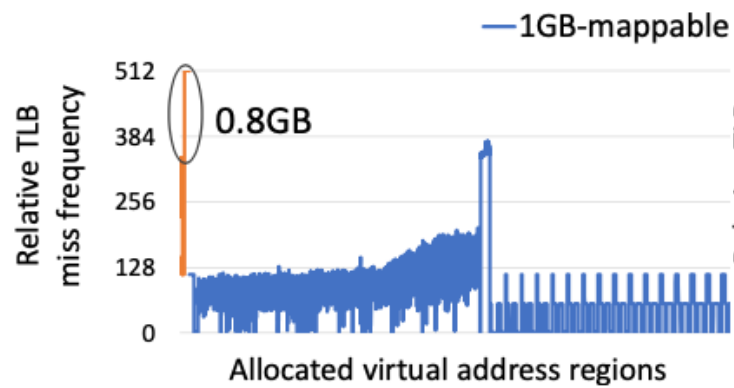} &  \includegraphics[width=0.5\columnwidth]{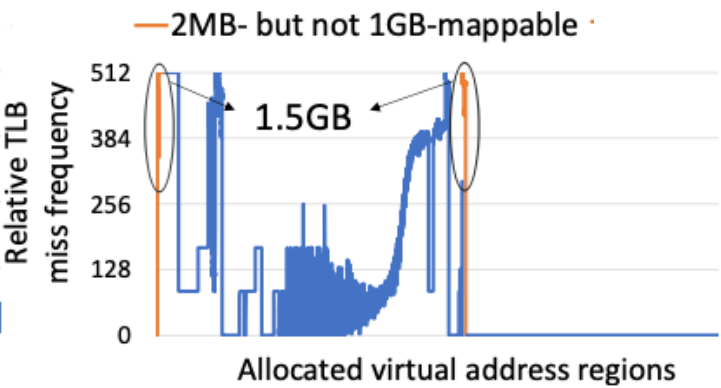} \\
    (a) Graph500 & (b) SVM \\
\end{tabular}
    \caption{Relative TLB-miss frequency.}
    \vs{-1.75em}
    \label{fig:access_maps}
\end{figure}

\subsection{Importance of using all large page sizes}
\label{subsec:2MB_importance}
In the analysis so far, \textit{only one} of the large page sizes was deployed as is the norm in today's software. 
However, we find that using all large page sizes together can bring benefits that are not achievable using any one of them.
%Further, deploying only the largest page size can also be counter-productive -- we will show why. 
%While it is obvious why 1GB page may bring benefit over 2MB pages, it may not be immhttps://www.overleaf.com/project/5d43bd43471b193cf6ff8a13ediately obvious under what circumstances one may wish to use 2MB pages beyond 1GB pages.

A virtual address range is \textit{mappable} by a large page only if:  \circled{1} it is at least as long as that large page, and \circled{2} the starting address is aligned at the boundary of that page size.
All 1GB-mappable address ranges are, thus, mappable by 2MB pages but \textit{not} vice-versa.
When an application allocates, de-allocates, and re-allocates memory (e.g., \textsf{Graph500}), the virtual address space gets fragmented.
Consequently, an application's entire address space may not be mappable by the largest page size. 

We empirically find that often GBs of an application's virtual memory is 2MB-mappable but \textit{not} 1GB-mappable.
This depends on the application's memory allocation strategy -- whether the application pre-allocates memory in large chunks (low \textit{virtual memory} fragmentation) or incrementally allocate/de-allocate memory over time (high fragmentation).  

\ignore{
The difference between the two depends on an application's memory allocation strategy -- it is larger if the virtual address space gets fragmented due to an application's incremental allocation, de-allocation, re-allocation of memory (e.g., \texttt{graph500}) and/or due to non-large page aligned allocation requests.}
\ignore{
For example, if the application allocates most memory upfront in the execution and does not de-allocate/re-allocate (e.g., \texttt{GUPS}) or slowly over time (e.g., \texttt{Redis}) and if the size of allocations requests are large page size aligned (e.g., \texttt{GUPS}'s allocation are often not 1GB aligned as it adds small metadata). }

We wrote a kernel module to periodically scan an application's virtual address space to measure 1GB-mappable and 2MB-mappable address ranges. 
%Figure~\ref{fig:mappable} shows the results for two representative applications -- \texttt{Graph500} and \texttt{SVM}. 
\autoref{fig:mappable} shows the size of allocated virtual memory that is mappable with 2MB and 1GB over time for two representative applications -- \textsf{Graph500} and \textsf{SVM}. 
The x-axis represents the execution timeline (excluding initialization), and the y-axis is the virtual memory (in GB).  
%We exclude the initialization phase from the timeline.
The two lines in each graph show the amount of 1GB- and 2MB-mappable memory. 
We observe that several GBs of memory is mappable by 2MB pages but not by 1GB (the gap between the two lines). 
%We observe similar behavior in other applications, though it is not presented due to lack of space. 
If \textit{only} 1GB pages are used, these memory regions have to be mapped with 4KB pages while wasting 2MB TLB resources.

We then analyzed the importance of mapping the regions that are un-mappable by 1GB pages, with 2MB. 
We wrote another module to measure the relative (sampled) TLB miss frequencies to the addresses that are mappable by 2MB but not by 1GB and those by both. 
We periodically un-set the \textsf{access bits} in PTEs (4KB) and then tracks which ones get set again by the hardware, signifying a TLB miss.
\autoref{fig:access_maps} presents the measurement.  
The x-axis shows the allocated virtual address regions, and the y-axis shows the relative TLB miss frequencies to pages in those regions. 
We use different colors for 2MB-mappable but 1GB-unmappable, and 1GB-mappable addresses. 
We observe that the 1GB-unmappable regions witness frequent TLB misses. 
Particularly for \textsf{Graph500}, the spike in miss frequency on a relatively small 1GB-unmappable region (about 800MB) stands out (circled). 
Therefore, it is important to map these 1GB un-mappable address ranges with 2MB pages to reduce TLB misses.

Furthermore, it may not always be possible to map a 1GB-mappable address range with 1GB page due to unavailability of 1GB contiguous physical memory.
However, 2MB contiguous physical memory regions are more easily available.
In short, it is \textit{important to utilize all page sizes available}. 

%\newtext{
We also measured the usefulness of 1GB pages to Linux kernel itself. 
Linux kernel \textit{direct} maps entire physical memory with the largest page size. 
Using a set of OS intensive workloads, we found that 1GB pages improve performance by around $2$-$3\%$ over 2MB pages (detailed in the Appendix). 
\noindent\textbf{Summary of observations:} 
\textcircled{1} A set of niche but important memory-intensive applications speed up with 1GB pages over 2MB pages.
In contrast, 2MB pages almost universally benefit memory-intensive applications. 
\textcircled{2} Application-transparent allocation of 2MB pages brings benefits of 2MB pages without user dependency -- a capability that 1GB pages lack.
\textcircled{3} It is important to utilize \textit{all} large page sizes \textit{not only} the largest. 
%\end{tcolorbox}

%\textbf{\textcircled{3}} Under virtualized environments the relative importance of 1GB pages increase a bit as slightly more performance gain is observed due to the use of 1GB pages compared to native execution. 
%\textcircled{4}} Linux kernel is one of the more intensive users of 1GB pages. \ignore{We quantitatively find that there is non-negligible impact on kernel's performance due to use of 1GB pages (relative to 2MB pages) when applications heavily exercise kernel functionalities.}
\ignore{\textbf{\textcircled{5}} In certain cases, application and kernel can contend for same 1GB TLB entries leading to overall performance degradation for the application. }

\section{Trident: Dynamic allocation of all page sizes}
\label{sec:THP_1G}
We design and implement \trident{} in Linux to enable application-transparent dynamic allocation of all three page sizes on x86-64 processors. 
\trident{} minimizes TLB misses by mapping most of an application's address space with 1GB pages, failing which 2MB, and finally, 4KB pages are used. 
%It does so without user or programmer guidance.

\noindent
\textbf{Challenges:} While the dynamic allocation of 2MB pages is not new, that for 1GB pages gives rise to many new challenges. 
First, \trident{} needs to ensure a steady supply of free contiguous 1GB physical memory chunks even in the presence of fragmentation.
%Otherwise, 1GB pages cannot be deployed. 
We found that Linux's sequential scanning based compaction for creating 2MB chunks is not scalable to 1GB due to excessive data copying. 
%We need a \textit{new} way .

Linux tracks only up to 4MB free physical memory chunks. 
However, the dynamic allocation of 1GB pages would require maintaining free memory upto 1GB granularity. 
Besides, allocating a 1GB page during a page fault is much slower than that for a 2MB or 4KB page due to the latency of zeroing entire 1GB memory. 
Low-latency 1GB page faults are necessary for an aggressive deployment of 1GB pages. 
%This impedes allocation of 1GB pages during a fault, a technique widely used for aggressive allocation of large pages (e.g., in \thp{}).
Finally, \trident{} should map a virtual address range with the largest \textit{large page} size deployable at that given time. 
%A page size is deployable if an address range is mappable by that page size and free physical memory chunk of requisite size is available.  
It should then periodically look for opportunities to promote address ranges mapped with a smaller large page(s) to a larger one wherever possible.
%Consequently, it should promote 2MB pages to 1GB pages when possible.  
%Finally, 1GB contiguous physical memory and 1GB TLB entries are both scarce resources.
%Otherwise, overall system performance could degrade.

\subsection{Design and implementation}
\label{subsec:design}
%We now describe our design and implementation of \trident{}.
%In the process, we will detail our approach for addressing the aforementioned challenges along with the associated quantitative analysis that guided our design decisions.

At a high-level, \trident{} modifies four major parts of Linux. 
\textcircled{1} It enhances Linux to track up to 1GB free physical memory chunks.
\textcircled{2} It updates the page fault handler to allocate 1GB page on a fault when possible and fall back to smaller pages if needed. 
\textcircled{3} \trident{} extends \thp{}'s \textsf{khugepaged} background thread to promote virtual address ranges to be re-mapped (promoted) to 1GB pages when possible.  
\textcircled{4}  \trident{} employs a novel \smartcompaction{} technique to ensure a steady supply of 1GB physical memory chunks at low overhead.

\subsubsection{Managing 1GB physical memory chunks.}
Linux's buddy allocator keeps an array of free lists of physical memory chunks of sizes 4KB up to 4MB in the power of 2~\cite{buddy}. 
%Thus the buddy allocator has a total of eleven lists.
%A node in a list tracks the starting physical address of a free memory chunk of a given size.
When free memory is needed, the buddy allocator provides a memory chunk from one of its lists based on the request size.
Freed physical memory is returned to the buddy, and coalesced with neighboring free memory chunks to create larger ones. 
Unfortunately, the buddy only keeps track of regions up to 4MB. 
We thus extended it to include separate lists for tracking up to 1GB memory chunks.  

\ignore{
\begin{figure}[tb]
\centering
\includegraphics[scale=0.40]{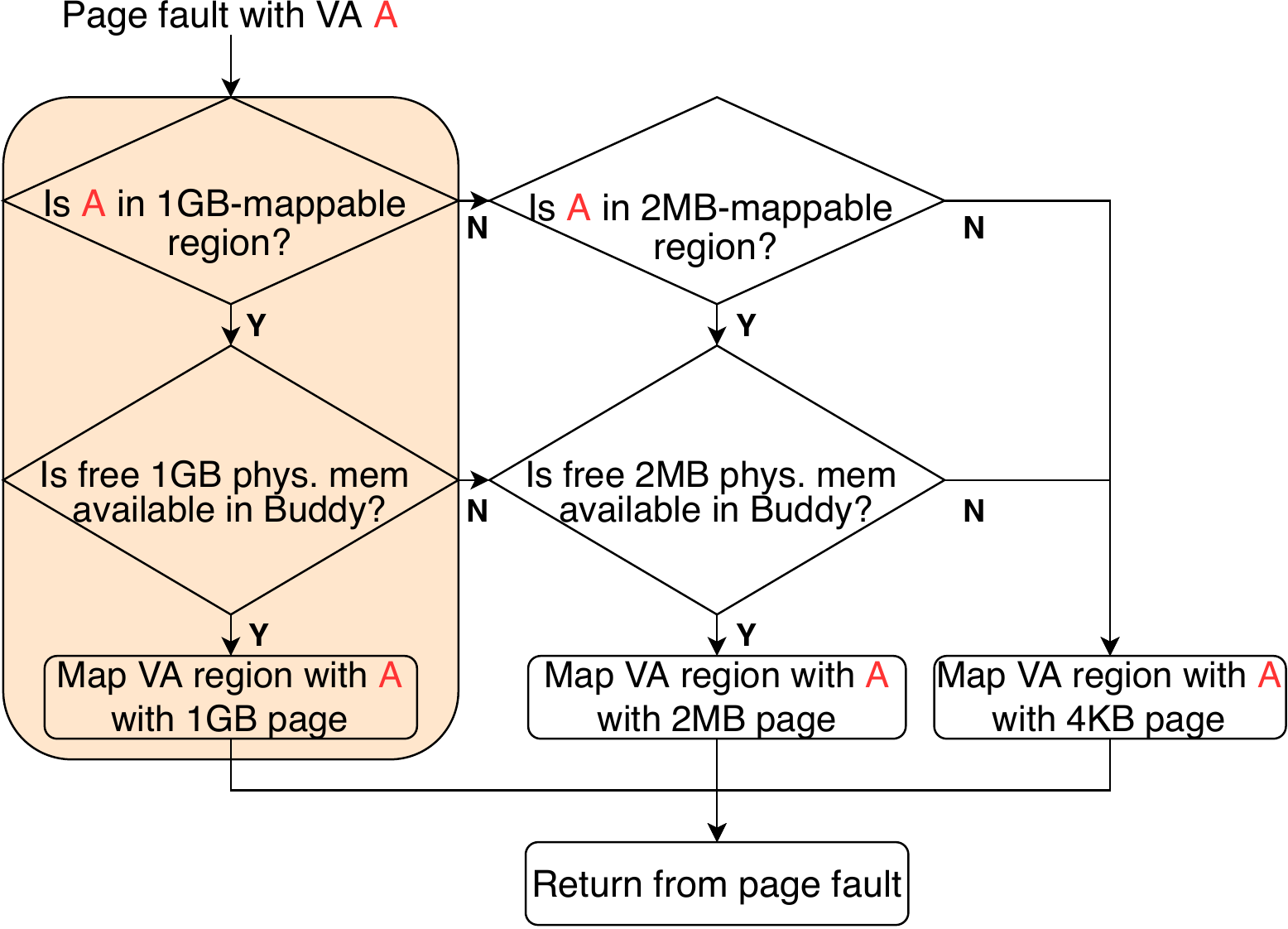}
\caption{Trident's allocation of page sizes during page fault. Trident-specific enhancement to THP is in shade.}
\label{fig:flowchart_pf}
\end{figure}
}

\subsubsection{Allocating large pages during page fault.}
Like \thp{}, \trident{} allocates large pages either \textcircled{1} during a page fault (e.g., when a process accesses a virtual address for the first time) or \textcircled{2} later during attempts to promote an address range to a large page. 
We here detail the former.

If the faulting virtual address falls in a 1GB-mappable address range, then \trident{} attempts to map it with a 1GB page. 
If it fails, \trident{} attempts to map the address with a 2MB page, and on failure, with 4KB. 
If the faulting address falls in a region that is 2MB-mappable but not 1GB-mappable \trident{} tries to map it with 2MB. 

\noindent
\textbf{Asynchronous zero-fill:} 
A 1GB page fault takes around $400$ milli-seconds; compare that to $850$ $\mu$seconds for 2MB. The additional latency is due to zero-filling of 1GB memory  instead of 2MB~\footnote{Zero fill ensures that no leftover data is leaked and cannot be avoided.}.  We instead employ asynchronous zero-fill to speed up 1GB faults.
A kernel thread periodically zero-fills free 1GB regions and \trident{} allocates an zero-filled region, if available.
This reduces the average 1GB fault latency from $400$ milli-seconds to $2.7$ milli-seconds. 
While prior works~\cite{panwar:asplos:2019, async-zero} explored asynchronous zero-fill for 2MB pages, we find it to be a necessity for 1GB pages in latency-critical workloads. 
For example, the boot time of a 70GB virtual machine dropped from $25$ seconds to $13$ seconds with asynchronous zero-fill.

\autoref{tab:1GB_alloctaed} shows the portion of applications' memory footprints mapped by 1GB and 2MB pages under \trident{}'s various dynamic allocation mechanisms that we will discuss in this section. 
The first data column shows applications' memory footprint. 
The first set of sub-columns capture the behavior with \textit{un-fragmented} physical memory while the next set represents that under \textit{fragmentation}. 
Physical memory is said to be fragmented if free memory is scattered in small holes and thus, non-contiguous. Typically, physical memory is un-fragmented \textit{only} if the system is freshly booted and/or there is little memory usage. 
But, the memory gets quickly fragmented as applications/OS allocate and deallocate memory. 

The sub-columns under \textsf{un-fragmented} shows that the page fault handler alone (\textsf{Page-fault only}) is able to map a large fraction of application's memory with 1GB pages for three out of eight applications (\textsf{XSBench, GUPS, Graph500}). 
If an application pre-allocates its memory in large chunks, then the fault handler would often find the faulting address to be in a 1GB-mappable region, and map it with a 1GB page. 
However, \textsf{Redis} and \textsf{Memcached} progressively allocate memory while inserting key-value pairs. 
Thus, the fault handler could map a small portion of its memory with 1GB pages. 
\textsf{SVM, Btree, Canneal} do not pre-allocate their entire memory needs, either.

The story is different if the physical memory is fragmented.
Even if the fault handler finds a 1GB-mappable address range, it is unlikely to find a free 1GB physical memory chunk. 
Thus, it would often fall back to 2MB or 4KB pages. 
This is evident from data in sub-columns for ``Page-fault only'' under fragmentation (\autoref{tab:1GB_alloctaed}) as only a few 1GB pages are allocated.

\begin{figure}[tb!]
\centering
\includegraphics[scale=0.45]{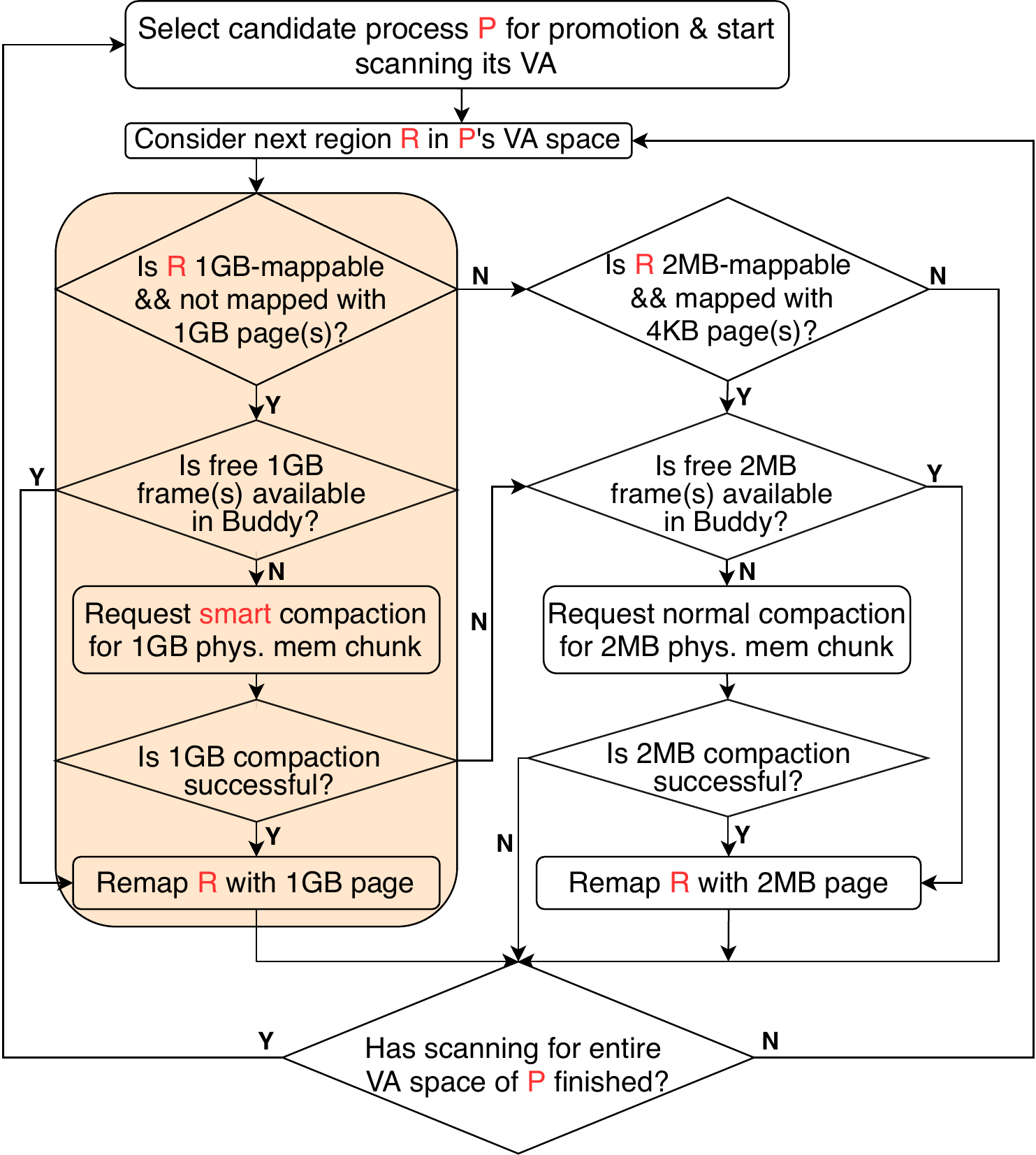}
\caption{Trident's large-page promotion algorithm.}
\vs{-1.5em}
\label{fig:flowchart_promotion}
\end{figure}

\begin{table*}[tb]
\caption{Comparison of 1GB and 2MB pages allocated via different mechanisms employed in Trident.}
\centering
\footnotesize
\begin{tabular}{@{}|l||l||l|l|l|l|l|l||l|l|l|l|l|l|@{}}
\hline
\multirow{3}{*}{} & \multirow{3}{*}{\begin{tabular}[c]{@{}l@{}}Memory\\ footprint \\ (in GB)\end{tabular}} & \multicolumn{6}{c||}{\textbf{Un-fragmented (all data in GB)}}   & \multicolumn{6}{c|}{\textbf{Fragmented (all data in GB)}}  \\ \cline{3-14}  & & \multicolumn{2}{l|}{\begin{tabular}[c]{@{}l@{}}Page-fault\\ only\end{tabular}} & \multicolumn{2}{l|}{\begin{tabular}[c]{@{}l@{}}Promotion \\ Normal compaction\end{tabular}} & \multicolumn{2}{l||}{\begin{tabular}[c]{@{}l@{}}Promotion \\ Smart  compaction\end{tabular}} & \multicolumn{2}{l|}{\begin{tabular}[c]{@{}l@{}}Page-fault\\ only\end{tabular}} & \multicolumn{2}{l|}{\begin{tabular}[c]{@{}l@{}}Promotion \\ Normal compaction\end{tabular}} & \multicolumn{2}{l|}{\begin{tabular}[c]{@{}l@{}}Promotion\\ Smart compaction\end{tabular}} \\ \cline{3-14} 
&  & 1GB  & 2MB  & 1GB  & 2MB  & 1GB  & 2MB  & 1GB  & 2MB  & 1GB   & 2MB  & 1GB  & 2MB \\ \hline \hline
XSBench  & 117 & 114 & 2.94  & 116  & 1.2 & 116 & 1.2  & 6  & 5.3 & 79 & 38.1 & 80 & 37.1 \\ \hline
GUPS & 32 & 31 & 1 & 31 & 1 & 31 & 1 &  9 & 2.5 & 31 & 1 & 31 &  1 \\ \hline
SVM & 68.5 & 54 & 14.3 & 65 & 3.5  & 65 & 3.5  &  6 & 5  & 53 & 12.2 & 54 & 9.9 \\ \hline
Redis & 44 & 0 & 0.5 & 39 & 3.4 &  39  &  3.4  &   0  &   0   &  25  &  10.3  &  28  & 14.3 \\ \hline
\highlight{Btree} & \highlight{25} & \highlight{0} & \highlight{16.7} & \highlight{16} & \highlight{5.8} & \highlight{16} & \highlight{5.8} &  \highlight{0} & \highlight{11.7} & \highlight{8} & \highlight{12.73} &  \highlight{12} & \highlight{8.91} \\ \hline
Graph500 & 63.5 & 59 & 4.01 & 60 & 3.35 & 60 & 3.35 & 5 & 5.8 & 37 & 24.2 & 38 & 23.6 \\ \hline
Memcached & 137 & 16 & 121 & 121 & 16 & 121 & 16 & 9 & 60 & 12 & 55 & 16 & 60 \\ \hline
Canneal & 32 & 8 & 1 & 30 & 2 & 30 & 2 & 6 & 1 & 6 & 21 & 8 & 22 \\ \hline
\end{tabular}
\label{tab:1GB_alloctaed}
\end{table*}

\subsubsection{Large page promotion.}
If an application does not pre-allocate memory or the physical memory is fragmented, it becomes important to later re-map (promote) address ranges with larger pages, when possible. 
\trident{} extends \thp{}'s \textsf{khugepaged} thread to promote to both 1GB and 2MB pages.

%Re-mapping  (promoting) a virtual address region currently mapped with smaller page size with larger page size larger 
%\trident{} attempts to re-map (promote) 1GB-mappable address range that is currently mapped with smaller page sizes, with 1GB pages. 
%Failing which, it attempts to map with 2MB (if not already mapped with 2MB). 
%\trident{} extends \thp{}'s  \texttt{khugepaged}, the kernel thread that traditionally promote mappings for both 2MB and 1GB.
%Promotion is key for applications that do not pre-allocate memory and/or if the physical memory is fragmented.

\autoref{fig:flowchart_promotion} shows a flowchart of \trident{}'s page promotion algorithm (changes to \thp{} are shaded).
\textsf{khugepaged} first selects a candidate process whose memory for promotion and sequentially scans its virtual address space. 
During scanning, \trident{} looks for a 1GB-mappable virtual address ranges that are mapped with smaller pages.
Subsequently, it looks for 2MB-mappable regions mapped with 4KB. 
If a candidate 1GB-mappable range is found, \textsf{khugepaged} requests the buddy allocator for a free 1GB physical memory chunk. 
If a 1GB chunk is unavailable, \textsf{khugepaged} requests compaction of the physical memory to create one. 
\trident{} extends \thp{}'s compaction functionality to create a 1GB physical memory chunk (will be detailed shortly). 
If the compaction fails, it attempts to map it with 2MB pages (if not already mapped with 2MB).
\trident{}'s policy of preferring 1GB pages but falling back to 2MB pages, makes the most out of TLB resources.

\autoref{tab:1GB_alloctaed}'s sub-columns under ``normal compaction'' shows the number of 1GB and 2MB pages allocated when the above-mentioned promotion policy is applied along with the page fault handler (under both un-fragmented and fragmented memory). 
For example, in the un-fragmented case, \textsf{khugepaged} is able to promote about 39GB of memory using 1GB pages for \textsf{Redis}, when the fault handler alone failed to allocate even a single 1GB page.  
\textsf{SVM, Canneal} also enjoyed many more 1GB pages due to page promotion.

When the physical memory is fragmented, page promotion helps applications get some 1GB pages, although slightly smaller in number compared to the un-fragmented case. 
For example, 1GB pages allocated to \textsf{SVM} drops from $65$ to $53$. 
This is expected; free 1GB memory chunks are  scarce even after compaction.

Overheads of compaction for 1GB, however, can negate benefits of 1GB pages.
Creating even a \textit{single} 1GB chunk often requires significant memory copying. 
Copying data creates contention in memory controllers and pollutes caches.
%These impact application performance. 
It also requires scanning larger portions of physical memory during compaction.
%This also shows up as large CPU utilization by the \texttt{khugepaged} thread. 
The applications threads could get a smaller fraction of CPU cycles as they can contend with kernel threads performing compaction.
In short, it is necessary to reduce the cost of compaction for 1GB pages to harness its full benefit.

\begin{figure}[t!]
\centering
%\includegraphics[width=2\columnwidth]{./figs/smart_compact.pdf}\\
%\vspace{-1em}
\includegraphics[scale=0.35]{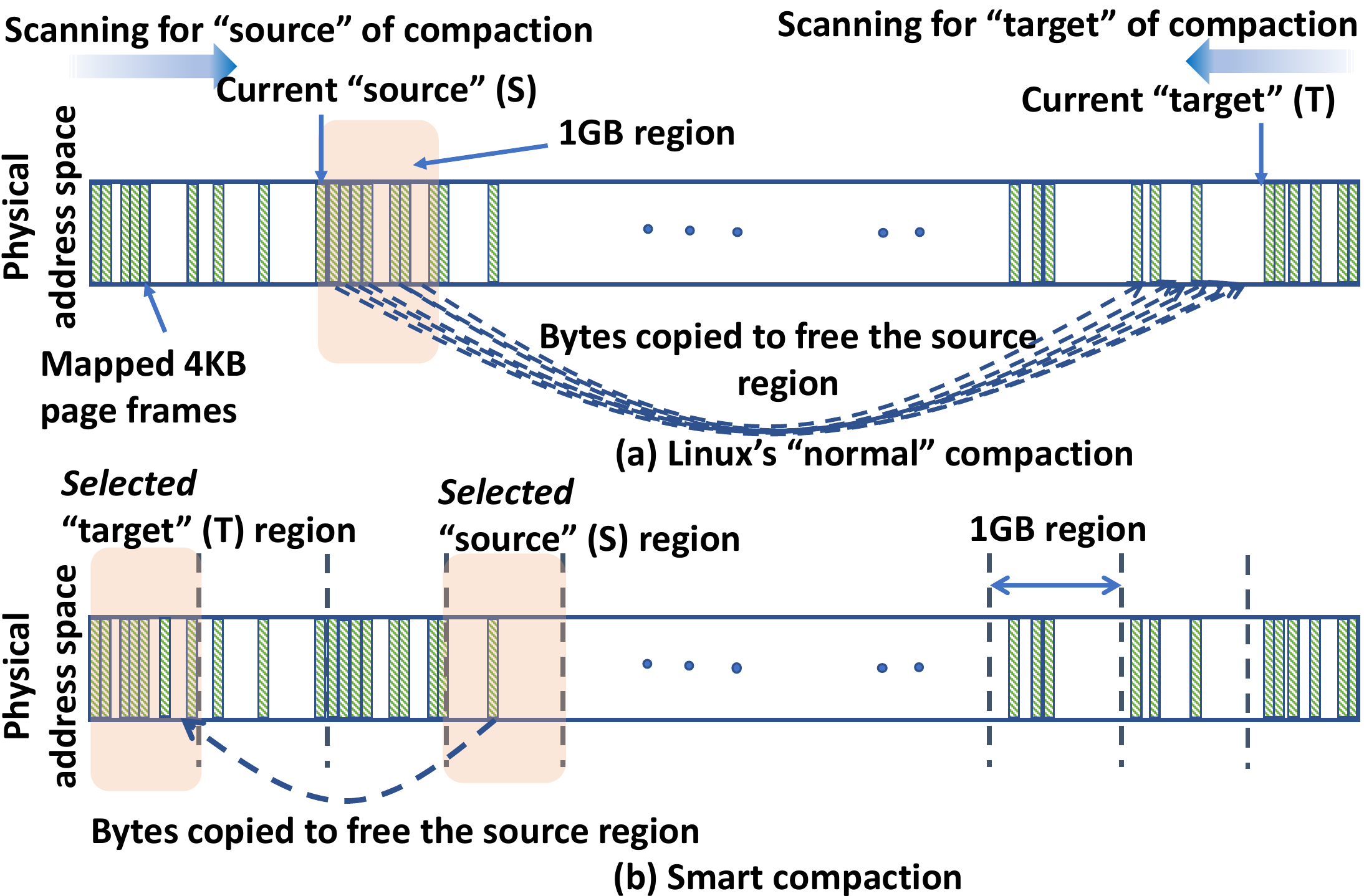} \\
\caption{Comparison between Linux's (traditional) compaction of physical memory and Smart compaction.}
\vs{-1.5em}
\label{fig:smart_compact}
\end{figure}

\noindent
\textbf{Smart compaction:}
We, thus, propose a new compaction technique, called \smartcompaction{}, to reduce the cost of 1GB compaction while creating enough 1GB physical memory chunks.
The primary goal is to reduce the number of bytes copied.
This directly reduces the cost of compaction. 
\ignore{
A secondary but related goal is to reduce the amount of scanning that is required to identify which 1GB region to compact and which bytes to copy to where. 
This could help in reducing CPU cycles spent in compaction.}

\autoref{fig:smart_compact} illustrates the difference between \textsf{normal compaction} as employed in Linux today and the \smartcompaction{} employed in \trident{}. 
\autoref{fig:smart_compact}(a) shows the working of the normal compaction. 
On a compaction request, the \textsf{khugepaged} thread starts sequentially scanning physical memory from where it left last time it attempted to compact (remembered in \textit{source} pointer). 
Scanning starts from the low to high physical address. 
As it finds an occupied physical page frame (4KB) it copies its contents to a free page frame found by scanning in the opposite direction from the \textit{target} pointer.
This continues until a free memory chunk of the desired size (e.g., 2MB) is created, or the entire memory is scanned without success.  

We observe that this strategy is agnostic to how full or empty a physical memory region is.  
Consequently, this leads to redundant copying.
Let us consider the example in \autoref{fig:smart_compact}(a). 
The 1GB region starting at address \texttt{S} is mostly occupied and has only $256$ free page frames (4KB). 
Thus, to free that 1GB region, Linux would require to copy $999$MB of data ($512\times512$ - $256$ 4KB pages). Instead, if a \textit{mostly} free region was freed, then the number of bytes copied would be much smaller.
While such sub-optimal compaction could be fine for 2MB, it is \textit{not} so for 1GB, as the data copying increases with size. 

\ignore{
If the chosen target region \texttt{T} has enough free base pages then data is copied there. 
Otherwise, another target region \texttt{T} is chosen to copy the left over. 
Inherently, this is sub-optimal.}

Moreover, if the scan encounters a page frame with unmovable contents (e.g., inodes, DMA buffers)~\cite{panwar:asplos:2019}, then all copying so far for a region, is wasted. 
A free chunk cannot have any unmovable contents.
The probability of encountering unmovable contents is much more for a 1GB region. 

\ignore{
the compaction of a selected \texttt{source} region (i.e., \texttt{S}) can fail for multiple reasons -- \textbf{\textcircled{1}} if there is even a \textit{single} base page within the source region that is pinned (thus, unmovable), or  \textbf{\textcircled{2}} if there is even a \textit{single} base page in \texttt{S} containing unmovable kernel data (e.g., \todo{}) or  \textbf{\textcircled{3}} if there is more than allowed amount of contention for copying pages. 
The probability of any of this condition of being true goes up with the number of allocated base pages within the \texttt{source} region. 
If the compaction fails to free the \texttt{source} region then the work (e.g., data copying) already done is wasted. }

To address these shortcomings, we propose \textsf{smart compaction}.
The key idea is to divide the physical memory into 1GB regions and \textit{select} (\textit{not} scan for) a region with the least number of occupied page frames for freeing (i.e., the \textsf{source} of copying).
Similarly, a region with the most number of occupied page frames is preferred as the \textsf{target} for copying. 
This strategy minimizes data copy. 
We also track if a given 1GB region contains any unmovable contents. 
We avoid selecting regions with unmovable content for freeing (i.e, source).
This eliminates unnecessary data copying in futile compaction attempts.
%}
\ignore{
At the same time, it also reduces the chances of failure (and thus, wasted work for compaction) due to an unmovable page.
This is because larger the number of page frames whose contents to be moved higher is the chance that one of those page frames would contain unmovable data.  }
\ignore{The probability of failure due to excess CPU contention due to copying of data also reduces.
Less number of failures in compaction, in turn, reduces the amount of wasted work. }

\begin{figure}[tb!]
\centering
\includegraphics[scale=0.4]{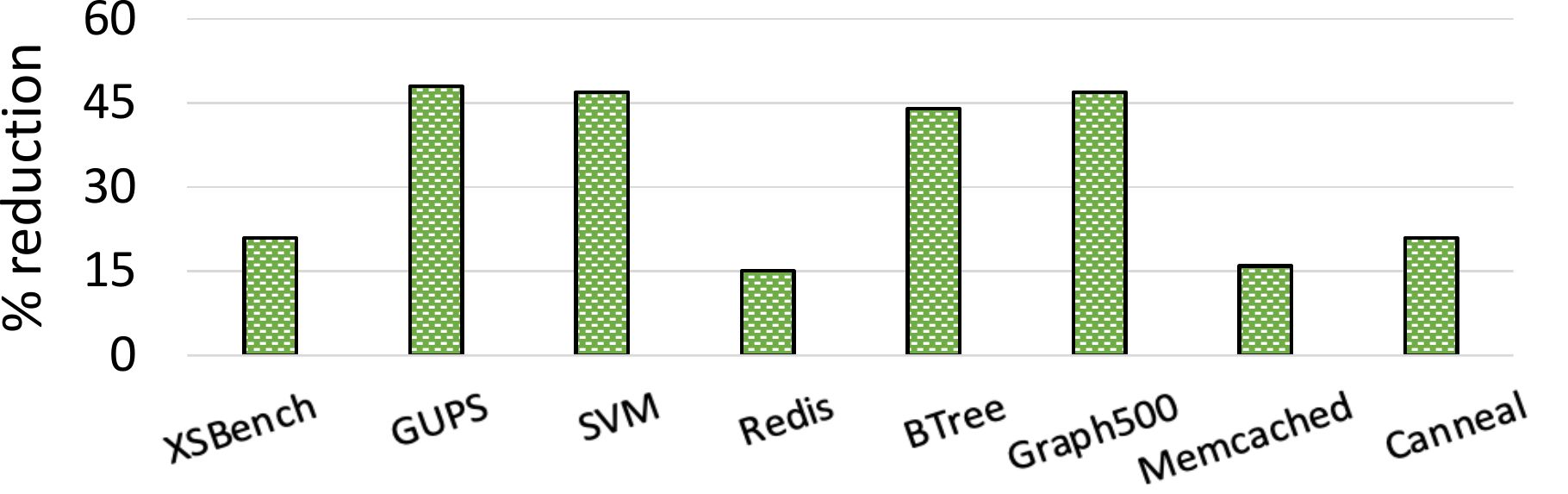}
\vs{-0.5em}
\caption{Reduction in bytes copied by smart compaction.}
\vs{-1.5em}
\label{fig:bytes_copied}
\end{figure}

\begin{figure*}[t]
    \centering
    \includegraphics[scale=1]{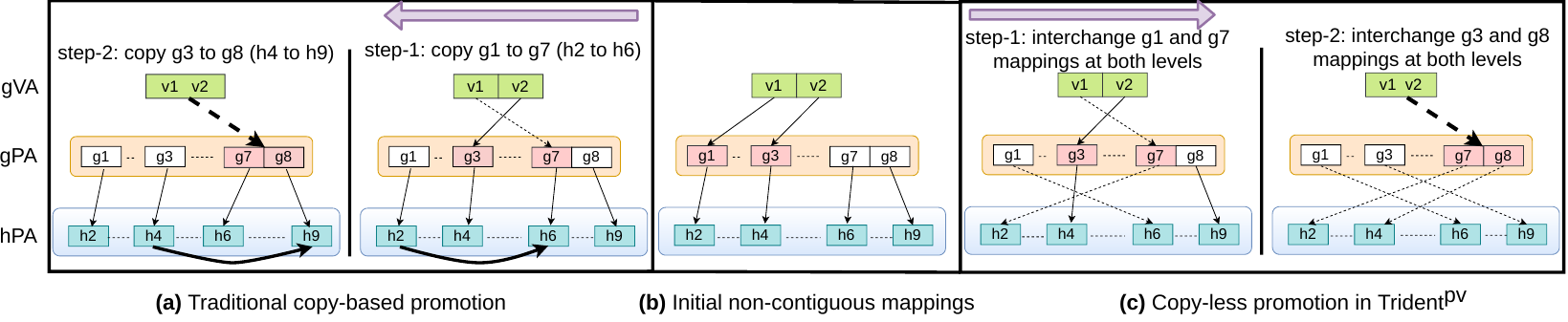}
    \vs{-0.5em}
     \caption{Traditional copy-based vs. \tridentpv{}'s copy-less page promotion.}
     \vs{-1.5em}
    \label{fig:trident_pv:design}
\end{figure*}

To implement the above idea, we first introduced two counters for each 1GB physical memory regions.
One counter tracks the number of free page frames, and the other one tracks the number of unmovable pages within a region. 
Whenever a page is returned to the buddy allocator (i.e., freed), we increment the counter for free frames of the encompassing 1GB region.
Further, we decrement the counter for unmovable pages if the freed page frame(s) contained unmovable data.
Whenever a page frame(s) is allocated from the buddy allocator the free counter for the encompassing region is decremented.
We increment its unmovable page counter if the allocated page frame(s) would contain unmovable data (e.g., requested for allocating kernel data structures).
Note that a 1GB region can also have a 2MB page allocated within it. 
We treat it as 512 base pages for ease of keeping statistics. 

As depicted in \autoref{fig:smart_compact}(b) the smart compaction starts by selecting a 1GB region with largest number of free page frames and without any unmovable pages as the \textsf{source(S)}. 
It then selects a target region (\textsf{T}) to move the contents of occupied page frames in the source.
The region with the least number of free page frames is selected as the target. 
It can happen that \textsf{T} may not have enough free frames to accommodate all of \textsf{S}'s page frames. 
If so, a region with next least number of free frames is selected to accommodate leftovers (and, so on). 

The sub-columns for \smartcompaction{} in \autoref{tab:1GB_alloctaed} shows the number of 1GB and 2MB pages that were allocated under un-fragmented and fragmented physical memory. 
We observe that the number of 1GB pages allocated to each application is the same as that under the normal compaction in the un-fragmented case. 
Under fragmentation, smart compaction typically provides even more 1GB page. 
This is because the smart compaction always selects a 1GB region that is easiest to free, and thus, compaction succeeds more often.

\autoref{fig:bytes_copied} shows the percentage reduction in the number of bytes copied with smart compaction over normal compaction. 
This measurement is performed when physical memory is fragmented as otherwise compaction is not required. 
We observe that \smartcompaction{} often reduces the number of bytes copied by more than half and up to 85\%. 
This demonstrates that \smartcompaction{} performs less work to create the same or more number of 1GB chunks. 
Only for \textsf{XSBench}, the improvement is less. 
\textsf{XSBench} uses a large fraction of total memory in the system and thus, even the ideal compaction algorithm would not be able to avoid data copy under fragmentation. 
All compaction algorithm will behave the same if the physical memory is fragmented, and an application needs all memory.

\noindent
\textbf{Summary:} \trident{} incorporates the modified page fault handler, the modified promotion logic (\autoref{fig:flowchart_promotion}), and \smartcompaction{} (\autoref{fig:smart_compact}) to dynamically allocate 1GB, 2MB or 4KB pages, as deemed suitable.

\section{\tridentpv{}: Paravirtualizing Trident}
\label{sec:trident_pv}
Under virtualization, \trident{} can be deployed both in the guest OS and in the hypervisor to bring benefits of dynamic allocation of all page sizes, including 1GB pages,  to both the levels of translation. 
We observe that it is possible to further optimize certain guest OS operations with paravirtualization.

The guest OS \textit{copies} contents of memory pages to \textbf{\textcircled{1}} to compact \textsf{gPA}s, and \textbf{\textcircled{2}} to promote address mapping between \textsf{gVA} and \textsf{gPA} to larger pages. 
While the cost of copying 4KB pages is not high, copying $2$MB pages in order to compact or promote to 1GB page is slow. 
We, however, observe that the effect of copying guest physical pages can be mimicked by simply altering the mapping between corresponding \textsf{gPA}s and \textsf{hPA}s.
This \textit{copy-less} approach quickens both compaction and 1GB page promotion in the guest but needs paravirtualization. 
We call this optional extension \tridentpv{}.

\ignore{
\tridentpv{} helps optimize two different use-cases: (1) promotion of large pages, (2) compaction of guest physical memory. The key mechanism behind copy-less construction of memory contiguity in \tridentpv{} is same in both the use-cases; for brevity, we explain the mechanism with the help of large page promotion (see \autoref{fig:trident_pv:design}).
}

For brevity, we explain the key idea behind \tridentpv{} with the help of large page promotion only (\autoref{fig:trident_pv:design}).
Let us assume that two contiguous guest virtual pages, \textsf{v1} and \textsf{v2}, are currently mapped to two non-contiguous smaller pages \textsf{g1} and \textsf{g3} in guest physical memory (\autoref{fig:trident_pv:design}(b)). 
For simplicity, we assume that a large page is double the size of a small page. 
To remap \textsf{gVA} encompassing \textsf{v1} and \textsf{v2} with a large page, the guest OS first copies their content to two contiguous guest target physical pages -- \textsf{g7} and \textsf{g8} and then updates the mapping between \textsf{gVA} and \textsf{gPA}. This traditional way of promoting large pages by copying contents is shown in \autoref{fig:trident_pv:design}(a).

\autoref{fig:trident_pv:design} (c) shows \tridentpv{}'s approach for page promotion without actual copy. 
Instead of copying \textsf{g1} to \textsf{g7}, the hypervisor exchanges the \textsf{gPA} to \textsf{hPA} mappings for \textsf{g1} and \textsf{g7}. 
After the exchange, \textsf{g1} would map to \textsf{h6} and \textsf{g7} to \textsf{h2}. Since, \textsf{h2} contains the data originally mapped by \textsf{g1}, this is same as copying \textsf{g1} to \textsf{g7}.  
Similarly, the hypervisor exchanges the \textsf{gPA} to \textsf{hPA} mappings for \textsf{g3} and \textsf{g8} to create the effect of copying \textsf{g3} to \textsf{g8}. Later, \textsf{gVA} encompassing \textsf{v1} and \textsf{v2} is mapped by the guest with a large page to contiguous
\textsf{gPA} encompassing \textsf{g7} and \textsf{g8}.

In this approach, the guest OS and the hypervisor need to coordinate for copy-less page promotion and, thus, the need for paravirtualization. Specifically, the guest OS supplies the hypervisor with a list of source and target guest physical pages via a hypercall. The hypervisor then updates the mapping from \textsf{gPA} to \textsf{hPA} in the manner explained above to create the effect of copying guest physical pages. Besides promotion, \tridentpv{} uses the same hypercall for compacting guest physical memory to create 1GB pages in the guest.

While promising, the cost of hypercall (\textasciitilde 300ns) to switch between guest and the hypervisors can, however, outweigh the benefits the copy-less promotion. We thus batch request for multiple page mapping exchanges in a single hypercall. Two pages are predefined for passing the list of page addresses to exchange between the guest and the hypervisor. One page contains source \gpa{}s (here, \textsf{g1} and \textsf{g3}) and the other contains the target \gpa{}s (here, \textsf{g7} and \textsf{g8}). In a single hypercall it is thus possible to request exchange for $512$ page addresses.  Thus, a single hypercall is sufficient to promote entire 1GB region in \gva{} mapped with 2MB pages. The hypercall returns after switching all the requested pages or logs any failure in the same shared page used for passing list of pages. On failure, the guest falls back to individually copy contents of pages.

We empirically found that promoting 2MB pages to a 1GB page in the guest takes \textasciitilde$600$ ms in the copy-based technique. 
Without batching, \tridentpv{} can promote the same in less than $30$ ms while batching reduces the time to \textasciitilde$500$ $\mu$s. 
Note that \tridentpv{}'s copy-less promotion is less useful for promoting 4KB pages to 2MB since the cost of copying 4KB pages is not significant. Hence, we employ copy-less promotion and compaction for 1GB pages only.
%}
\section{Evaluation}
\label{sec:eval}
We evaluate \trident{} to answer the following questions. 
\textcircled{1} Can \trident{} help improve performance of memory-intensive applications over Linux's default \thp{}, and over a recent work, called \hwk{}\cite{panwar:asplos:2019}?
\textcircled{2} How important is \trident{}'s use of all large page sizes?
\textcircled{3} What are the sources of performance improvement for \trident{}?
\textcircled{4} How does \trident{} perform under virtualization? 
\textcircled{5} Finally, how does \tridentpv{} impact page promotion/compaction the guest OS?

\begin{figure}[t]
    \centering
    \includegraphics[width=\columnwidth]{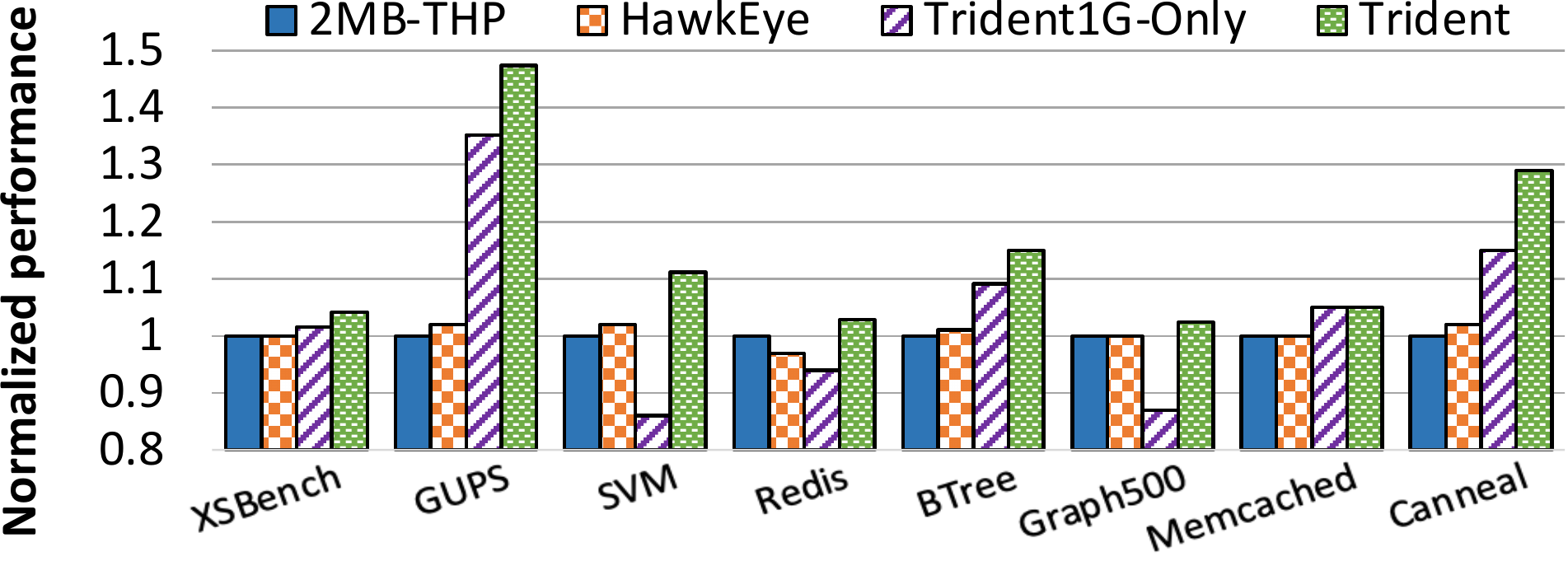}
     \vs{-1.5em}
     \caption{Performance under no fragmentation.}
     \vs{-1.5em}
    \label{fig:trident_perf_unfrag}
\end{figure}

\noindent
\textbf{Performance under un-fragmented physical memory:}
\autoref{fig:trident_perf_unfrag} shows the normalized performance for four configurations (higher is better) -- \circled{1} Linux's \thp{}, \circled{2} \hwk{}, \circled{3} \tridentG{} and \circled{4} \trident{}. \hwk{} is the most recent related work that improved upon \thp{} and other previous works (e.g.,~\cite{kwon:osdi:2016}).
It does so by efficiently allocating 2MB pages to memory regions that suffer the most from TLB misses~\cite{panwar:asplos:2019}. This enables us to compare \trident{}'s performance to current state-of-the-art. \tridentG{} denotes the configuration where \trident{} is disallowed to use 2MB pages. The difference between \tridentG{} and \trident{} highlights the importance of leveraging all large page sizes.

For each application, there are four bars in the cluster corresponding to four configurations. The height of each bar is normalized to the performance of the application under \thp{} (Linux's default configuration). Measurements in \autoref{fig:trident_perf_unfrag} were performed when the physical memory is un-fragmented.

First, we observe that \trident{} improves performance over Linux's \thp{} by 14\%, on average and up to 47\% for \textsf{GUPS}. Applications like \textsf{XSBench}, \textsf{SVM}, \textsf{Btree}, \textsf{Canneal} witnessed $4.1\%$, $11.2\%$, $15\%$ and $30\%$ performance improvements, respectively. Even if we exclude the micro-benchmark \textsf{GUPS}, performance improvement is $12\%$, on average, over \thp{}.

Next, we observe that \trident{} also outperforms \hwk{} by $14\%$, on average. This is expected due to the similarities between huge page management in \textsf{Linux} and \hwk{} in an un-fragmented system where both utilize 2MB pages aggressively to maximize performance.

\begin{figure}[t]
    \centering
    \includegraphics[width=\columnwidth]{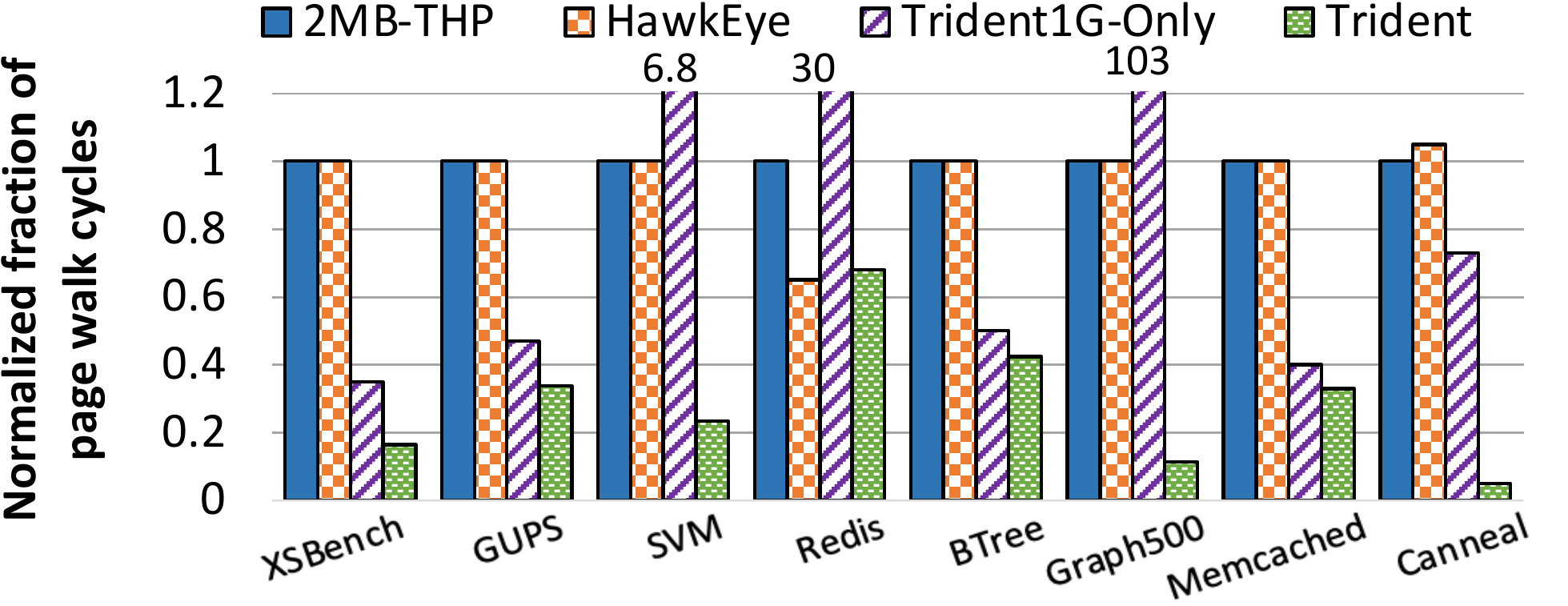}
    \vs{-1.5em}
     \caption{Fraction walk cycles under no fragmentation.}
     \vs{-0.5em}
    \label{fig:trident_walks_unfrag}
\end{figure}

Finally, there is a significant performance gap between \tridentG{} and \trident{}.
\tridentG{} loses performance even relative to \thp{} for several applications (e.g., \textsf{Graph500, SVM}). 
In hindsight, this is expected.
Our analysis in the Section~\ref{subsec:2MB_importance} revealed that these applications have significant portions of their virtual memory that is 2MB-mappable but not 1GB-mappable. 
Further, these portions also witness a relatively larger number of TLB misses. 
\tridentG{} is forced to map these 1GB-unmappable regions with base 4KB pages and thus have more translation overheads compared to \trident{} that could deploy 2MB pages. 
In the process, \tridentG{}'s benefits from using 1GB pages are more than negated by the overheads of mapping frequently accessed memory with 4KB pages. 

A keen reader will also observe that \tridentG{} performs much worse than  \textsf{1GB-Hugetlbfs} (\autoref{fig:unvirt_perf}, Section~\ref{subsec:analysis_univirt}) although neither of them uses 2MB pages. 
The reason is \textsf{libHugeTLBFS} maps the \textit{entire} chosen segment of an application's memory (here, heap) with 1GB pages, irrespective of the sizes of allocation requests.    For example, even if an application does a \textsf{malloc} for 12KB memory, \textsf{libHugeTLBFS} will map it with 1GB pages.   Thus, the question of 1GB-mappable vs. 2MB-mappable virtual address region does not arise under its static memory allocation mechanism. Since \tridentG{} enables dynamic memory allocation, it does not have any such luxury. Fortunately, \trident{} is able to more than makeup for it by utilizing all large page sizes while also retaining ease of programmability with dynamic allocation. 

\begin{figure}[t]
    \centering
    \includegraphics[width=\columnwidth]{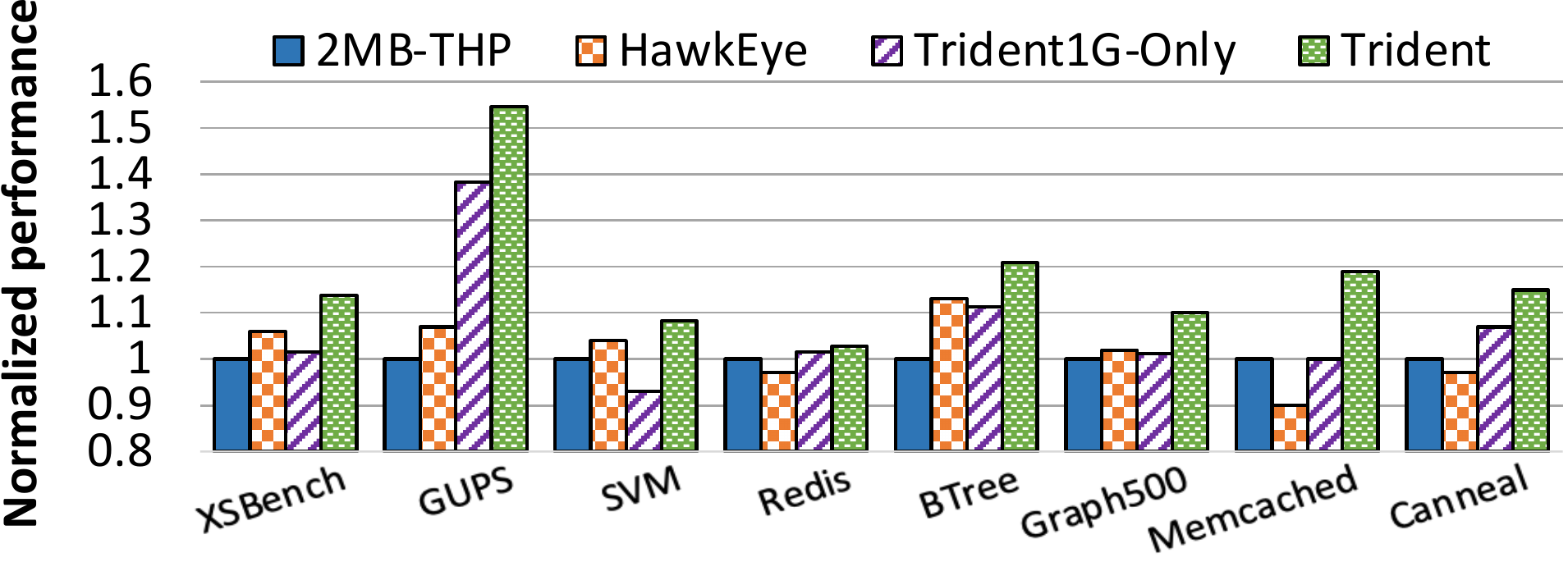}
    \vs{-1.5em}
     \caption{Performance under fragmentation.}
     \vs{-1.5em}
    \label{fig:trident_perf_frag}
\end{figure}

\begin{figure}[t]
    \centering
    \includegraphics[width=\columnwidth]{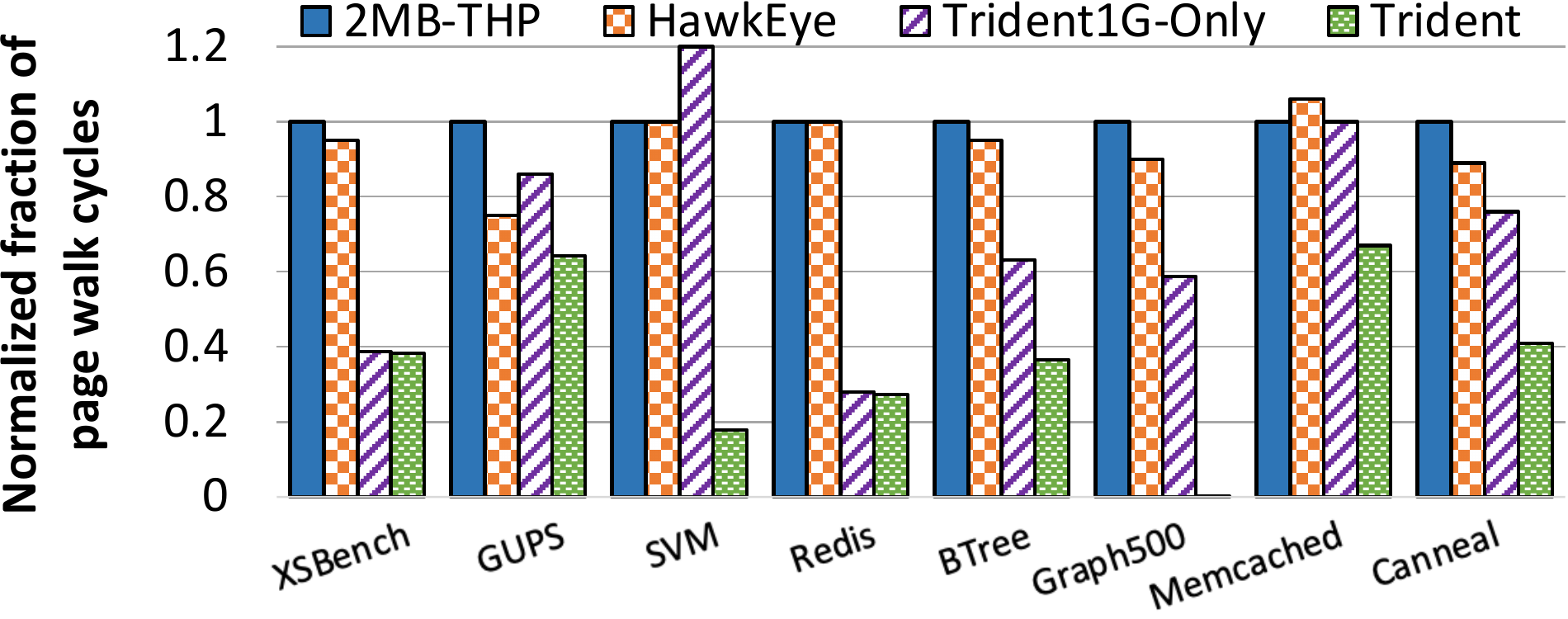}
    \vs{-1.5em}
     \caption{Fraction walk cycles under fragmentation.}
     \vs{-1em}
    \label{fig:trident_walks_frag}
\end{figure}

\noindent
\textbf{Performance under fragmented physical memory:}
Arguably, performance analysis under fragmented physical memory paints a more realistic execution scenario.
\autoref{fig:trident_perf_frag} shows the normalized performance under fragmented physical memory for the same four configurations as before. \trident{} speeds up applications even more under fragmentation. This is unsurprising since \trident{}'s smart compaction adds a further edge here. On average, it improves performance by $\textbf{18\%}$ over \thp{} and \textsf{GUPS} quickens by over $50\%$.
Even excluding \textsf{GUPS}, the improvement is $\textbf{13\%}$ over \thp{}.

\trident{} also outperforms \hwk{} in all cases. 
In some cases under fragmentation, \hwk{} performed worse than \thp{} (e.g., \textsf{Memcached}). 
Our discussions with the authors of \hwk{}
revealed that this might happen for large memory applications due to: \textcircled{1} CPU overhead of \textsf{kbinmanager} kernel thread that estimates relative TLB miss rates in \hwk{} and \textcircled{2} potential lock contention between \textsf{kbinmanager}, \textsf{khugepaged} and page-fault handler. 

\trident{} outperforms \tridentG{} by a good margin even under fragmentation.  However, note that in cases where \tridentG{} performed worse than \thp{} under un-fragmented physical memory, it performed almost equally under fragmentation. Here, \thp{} could deploy a lesser number of 2MB pages due to lack of contiguous physical memory. This, in turn, reduces the performance gap between \tridentG{} and \thp{}. The execution time for \thp{} also increases due to compaction
overhead, narrowing its edge further. 

\begin{table}[]
\caption{Percentage 1GB physical memory allocation failures}
\centering
\scalebox{0.72} {
\begin{tabular}{|l|l|l||l|l|l|} \hline
& \textbf{Page fault} & \textbf{Promotion} & & \textbf{Page fault} & \textbf{Promotion} \\ \hline
XSBench   & 94 & 32 & GUPS  & 71  & 0 \\ \hline
SVM       & 88 & 19 & Redis & NA  & 36 \\ \hline
Graph500  & 91   & 38   & Btree  & NA    & 25   \\ \hline
Memcached &  43  &  81  & Canneal   &  12  & 92  \\ \hline
\end{tabular}}
\label{tab:1Gfail}
\vs{-1.5em}
\end{table}

We also measured how often the fragmented physical memory prevents \trident{} from mapping an address with a 1GB page. \autoref{tab:1Gfail} shows the percentage of attempts to allocate a 1GB page that fails due to fragmentation.

There are ``NA" under page fault for \textsf{Redis} and \textsf{Btree} since the fault handler never attempts to allocate a 1GB page due to lack of 1GB-mappable virtual address range during faults. We observe that $71$-$94\%$ of 1GB page allocations fail due to lack of contiguous physical memory. Even during page promotion, 1GB allocations fail often. 
This further reinforces the need to utilize all large page sizes. Even if the largest page size cannot be used, a smaller large page (2MB) possibly be deployed.

\begin{table}[]
\vs{-1.5em}
\caption{Tail latency analysis for Redis}
\footnotesize
\centering
\begin{tabular}{|l||l|l|l|} \hline
 & 4KB-only & 2MB-THP & Trident \\ \hline \hline
No-fragmentation & 47.3 ms & 50.3 ms & 46.6 ms \\ \hline
Fragmentation & 53.3 ms & 53.3 ms & 52 ms \\ \hline
\end{tabular}
\label{tab:tail}
\end{table}

\noindent
\textbf{Impact on page walk cycles:}
\autoref{fig:trident_walks_unfrag} shows the normalized fraction of page walk cycles for \thp{}, \hwk{}, \tridentG{}, and \trident{} under un-fragmented memory.
\autoref{fig:trident_walks_frag} shows the same under fragmentation. The reductions in the fraction of walk cycles with \trident{} over \thp{} are significant -- $38$-$85\%$ under no fragmentation and $40$-$97\%$ under fragmentation. Across all configurations, we observe that relative improvement in performance correspond to relative reduction in page walk cycles.

\noindent 
\textbf{Impact on tail latency:}
Tail latency is an important for interactive applications (e.g., \textsf{Redis}) that need to abide by strict SLAs~\cite{kwon:osdi:2016}.
\autoref{tab:tail} reports 99 percentile latency of \textsf{Redis} with 4KB pages, \thp{}, and \trident{}, under both un-fragmented and fragmented memory.
\trident{} slightly improves tail latency, relative to both 4KB and \thp{}. 
\trident{} avoids long latency 1GB page faults by employing asynchronous zero-fill and 1GB pages reduces TLB misses in the critical path.

\ignore{
\noindent
\textbf{Memory bloat:}
A well-known drawback of using any large pages is the \textit{memory bloat}.
It happens when more physical memory is allocated to the application than it actually needs due to internal fragmentation.
Larger the page size more is the internal fragmentation and thus, bloat. 
\todo{Figure~\ref{fig:bloat}} shows the amount of physical memory allocated under \thp{}, \trident{} and \texttt{libHugeTLBFS-1GB}, normalized to that when entire memory is mapped with 4KB pages only (least possible bloat).  
We observe that except \texttt{Btree}, \trident{} adds almost zero bloat, while \texttt{libHugeTLBFS-1GB} allocates relatively more memory.
Only in case of \texttt{Btree}, \thp{} significantly bloats memory and \trident{} makes it even more sever.
The bloat due to \texttt{libHugeTLBFS-1GB} is even worse.
\texttt{Btree} has a very sparse access pattern where it uses only a small part of a page. Larger the page size the bloat increases. 

\trident{} does not attempt to reduce bloat as the key focus here is to make a case for using all large page sizes.
However, previously proposed techniques for reducing bloat, e.g., identifying internal fragmentation by scanning for all-zero page frames~\cite{panwar:asplos:2019}, can be readily applied over \trident{}, if desired. 
}

\begin{figure}[t]
    \centering
    \includegraphics[width=\columnwidth]{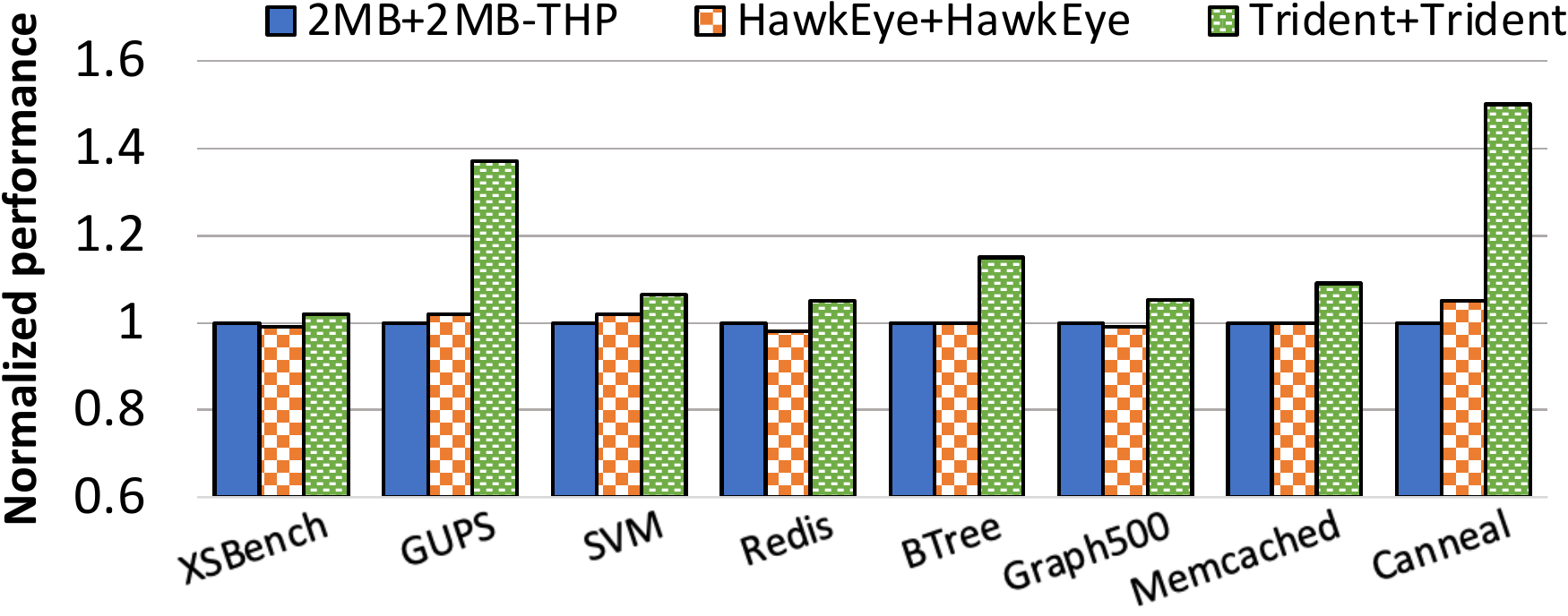}
    \vs{-1.5em}
     \caption{Performance under virtualization.}
     \vs{-1.5em}
    \label{fig:trident_virt}
\end{figure}

\begin{figure}[t]
    \centering
    \includegraphics[width=\columnwidth]{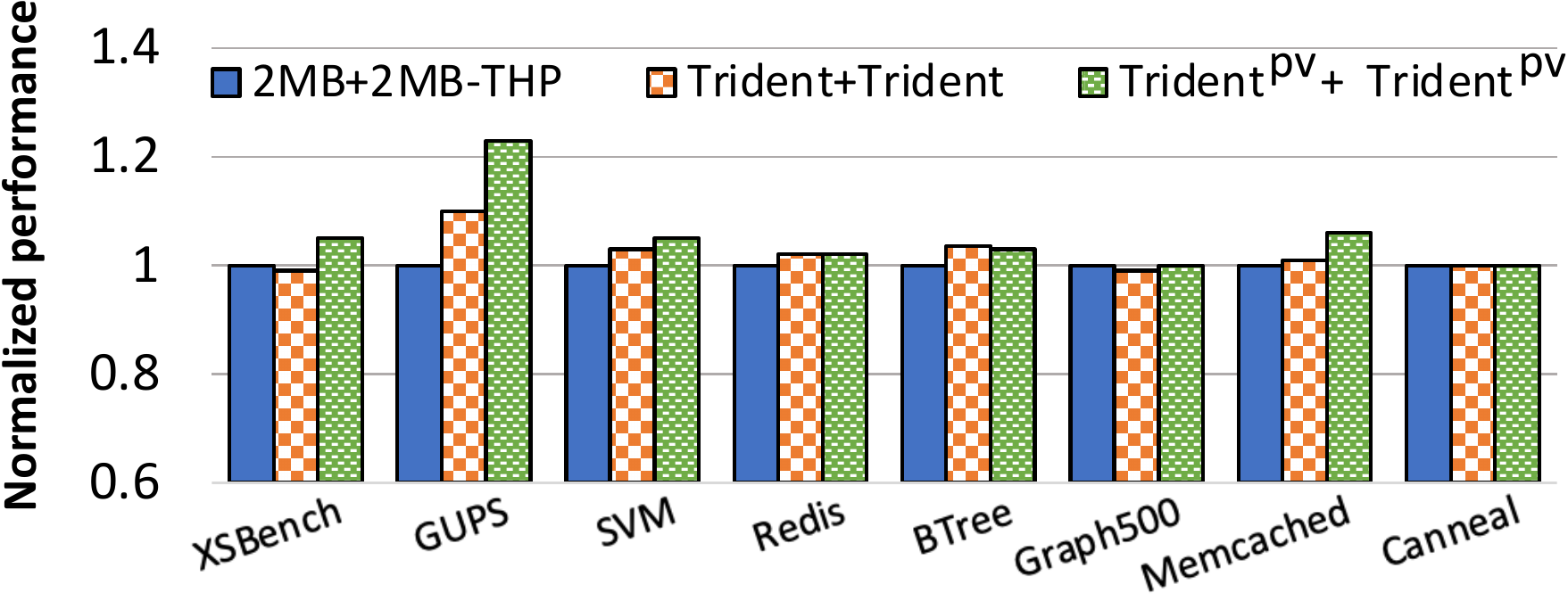}
    \vs{-1.5em}
     \caption{\tridentpv{}'s performance under fragmented gPA.}
     \vs{-1.5em}
    \label{fig:tridentpv_virt}
\end{figure}

\noindent
\textbf{Performance under virtualization: }
We measured the performance of applications running inside a virtual machine with \trident{} deployed both at the guest OS and at the hypervisor (KVM). We do not fragment memory. For comparison, we do the same with \hwk{} too. \autoref{fig:trident_virt} shows the speedups, normalized to \thp{} deployed in the guest OS and KVM. Under virtualization, \trident{} improves performance by $16\%$ on average, over \thp{} and by $15\%$ over \hwk{}. \textsf{Canneal} saw biggest improvement ($50\%$), but other applications also benefited significantly. For example, \textsf{SVM} and \textsf{Graph500} witnessed  6\% improvement each.

\noindent
\textbf{Performance with \tridentpv{}:} 
When the \textsf{gPA} gets fragmented over time, the guest OS must compact and promote pages with the help of \textsf{khugepaged} thread. However, a significant CPU usage by \textsf{khugepaged} in the guest OS could mean wasted vCPU time (cost) for a tenant in the cloud. In fact, Netflix reported how their deployments on Amazon EC2 can get adversely effected by high CPU utilization by \thp{}'s \textsf{khugepaged}~\cite{netflix:thp}. We, therefore, evaluate \tridentpv{} with fragmented \textsf{gPA} but limit \textsf{khugepaged}'s CPU utilization in the guest to maximum of $10\%$ of a single vCPU. This setup helps to find out whether \tridentpv{}'s faster copy-less promotion/compaction can be useful to use 1GB pages where CPU cycles are not free. 

\autoref{fig:tridentpv_virt} shows the performance of \trident{} and \tridentpv{} normalized to \thp{}. \tridentpv{} is more effective than \trident{} for \textsf{XSBench, GUPS, Memcached}, and \textsf{SVM} by 5\%, on average and by up to $10\%$. We also observe that \tridentpv{} does not always improve performance over \trident{}. Recall that \tridentpv{}'s hypercall-based copy-less approach is quicker than the copy-based approach only during promotion/compaction of 2MB pages to 1GB pages. Otherwise, the overhead of the hypercall and that of tracing and altering PTEs overshadows the benefits of avoiding copy. In applications such as \textsf{BTree, Graph500, Canneal}, 4KB pages are often promoted directly to 1GB pages without needing to go via 2MB pages limiting \tridentpv{}'s scope for improving their performance.

\noindent
%\newtext{
{\bf Memory bloat:} Large pages are well-known to increase memory footprint (bloat) due to internal fragmentation.
Larger the page size more is the bloat. 
\trident{} causes bloat in two out of eight workloads.
It adds 38GB and 13GB bloat for \textsf{Memcached} and \textsf{Btree} over \thp{}. 
We were able to recover the bloat by simply incorporating \hwk{}'s technique for dynamic detection and recovery of bloat by demoting large pages and de-deuplicating zero-filled small pages~\cite{panwar:asplos:2019}. 
However, we do not make any new contribution here and tradeoff between bloat and large pages is well explored~\cite{kwon:osdi:2016,panwar:asplos:2019}
%Incorporating this technique, we find that \trident{} can recover all memory bloat, albeit with a compromise on performance due to fewer 1GB pages.

%Balancing memory vs. performance tradeoff is an interesting design choice that is already detailed in HawkEye and Ingens papers; we do not contribute anything significant in this space.
%}

\noindent
\textbf{Summary:} We demonstrate that \trident{} significantly improves performance over \thp{} and state-of-art academic proposal~\cite{panwar:asplos:2019} under varied execution scenarios.
\section{Related work}
Address translation overheads is a topic of several recent research efforts spanning multiple fields ~\cite{doityourself,2dwalks,spgorman-characteristics,compiler-driven,largepagesvirtual,pomtlb,park:isca:2020}.
%True to the software-hardware design philosophy of the virtual memory, proposed solutions sometimes involve only the hardware, only the software or hardware and software both.
%We will first discuss previous works that require new hardware  and then those which are limited to only software modifications. 

\noindent
{\bf Proposals that require hardware support:} Hardware optimizations focus on reducing TLB misses or to accelerate page-walks. Multi-level TLBs, and multiple page sizes are found in today's commercial CPUs
\cite{hugepages-arch,colt-hugepages}.
Further, page walk caches are used to make page walks faster\cite{pwc,colt-mmu}.

Direct segments \cite{basu:isca:2013} can significantly reduce address translation overheads
through segmentation hardware. 
%OS-level challenges of memory fragmentation have also been
%considered in hardware designs:
Coalesced-Large-reach TLBs increase TLB coverage through contiguity-aware hints encoded in the page tables~\cite{colt}. This approach can also be combined with page walk caches and large pages ~\cite{colt-mmu, colt-hugepages, colt-hpca}. \textsf{POM-TLB} reduces page walk latency by servicing a TLB miss using a single memory lookup with a large in-memory TLB~\cite{pomtlb}.
\textsf{SpecTLB} speculatively provides address translation on a TLB miss by guessing virtual to physical address mappings~\cite{spec-tlb}.  
\textsf{ASAP} prefetches translations to reduce page-walk latency to that of a single memory lookup~\cite{asap:micro:2019}.
%It uses a base-plus-offset arithmetic that directly indexes into page-tables by ordering page-table pages to match the order of the virtual memory pages.}
It first orders page table pages to match that of the virtual memory pages and then uses a base-plus-offset arithmetic that directly indexes into the page tables. 
Large page support for non-contiguous physical memory has also been proposed~\cite{superpages-non-contiguous}. Tailor page sizes use whatever contiguity OS can afford to allocate~\cite{guvenilir:tps:2020}.
In contrast, \trident{} does not need new hardware; architectural support is complementary to \trident{}'s goal of fully utilizing large page support.

\noindent
{\bf Proposals with only software support:}
Software-only solutions mainly focus on better use of large pages. Navarro et. al., proposed a reservation-based large page promotion in FreeBSD as compared to proactive large page allocation in Linux~\cite{navarro}.
%This is a conservative approach to promote large pages compared to Linux's \thp{}. 
%Recently works like
\textsf{Ingens} proposes to mix \thp{}'s aggressive large page allocation with \textsf{FreeBSD}'s conservative approach to reduce memory bloat and latency while still leveraging large pages. 
\textsf{Illuminator} showed how unmovable kernel objects hinder compaction~\cite{panwar:asplos:2018}.
%\newtext{Quicksilver addresses software issues in large page management based on the sequence of events in the lifetime of a page; it aims to employ well-known mechanisms to handle each specific stage with the most suitable alternative~\cite{quicksilver:atc:2020}}.
Quicksilver uses hybrid strategies across different stages in the lifetime of a large page. Superficially, it employs aggressive allocation, hybrid preparation, relaxed mapping creation, on-demand mapping destruction and preemptive deallocation to achieve high performance and lower latency and bloat~\cite{quicksilver:atc:2020}.
\textsf{Carrefour-LP} showed how large pages can degrade performance in NUMA systems due to remote DRAM accesses and unbalanced traffic~\cite{carrefour-lp}. 
\trident{} is complementary to these works;  while these works focused on KBs and MBs-sized pages, \trident{} focuses on 1GB pages which brings its unique challenges for compaction and page promotion.  
Many insights from these works on 2MB pages are applicable to \trident{} too e.g.,  \textsf{HawkEye}'s fine-grained page promotion and \textsf{Ingens}'s adaptive approach of balancing trade-offs with large pages can be applied to \trident{} too.  

\textsf{Translation Ranger} proposed a new OS service to actively create contiguity~\cite{yan:isca:2019}.
%Authors suggest that the contiguity can be utilized by contiguity-aware hardware such as \textsf{coalesced-TLB}~\cite{colt}. 
%However, this service could degrades performance. 
Finally, a recent proposal allows in-place compaction of physical memory ~\cite{nvidia:1g:2019}.
This functionality, however, needs to be invoked via a special syscall and is unsuitable for applications that incrementally allocates memory (e.g., \textsf{Redis}). 
{\sf Mitosis}~\cite{mitosis} eliminates remote memory accesses due to page walks by replicating page tables on NUMA nodes; \trident{} can also help reduce the impact of remote accesses in page walks, as just a side effect.
In a concurrent work to ours, code patches to enhance \thp{} to support 1GB pages were recently posted on Linux mailing list\cite{thp-1g:2020}. 
This shows the Linux community's growing interest in 1GB pages. However, unlike these patches that simply aim to enable 1GB THP allocation, our work comprehensively shows the type of workloads that could benefit from 1GB pages, deals with physical memory fragmentation via smart compaction and virtualizes 1GB pages via copy-less 1GB page promotion and compaction in \tridentpv{}.

\ignore{
While most of these works focus on 2MB as the only large page size, \trident{} aims to utilize all three available page size support in current x86-64 processors.
Many of the techniques to improve the use of 2MB pages can be extended to 1GB pages too and thus \trident{} to a large extent is orthogonal to the aforementioned works. 
}

\ignore{
They proposed hardware-profiling based techniques to overcome
these problems.

\texttt{Ingens}  
 Mechanisms of dealing with
physical memory fragmentation and NUMA systems
are important but orthogonal problems,
and their proposed solutions can be adopted to improve the robustness of
HawkEye. Compiler or application hints
can also be used to assist OSs in prioritizing huge page
mapping for certain parts of the address space \cite{huge-huge,compiler-driven},
for which an interface is already provided by Linux through the {\tt madvise}
system call \cite{madvise}.

An alternative approach for supporting huge pages has also
been explored via {\tt libhugetlbfs} \cite{libhugetlbfs} where
the user is provided more control on huge page allocation. However,
such an approach requires manual
intervention for reserving huge pages in advance
and considers each application in isolation. Windows and OS X support huge
pages only through this reservation-based approach to avoid issues
associated with transparent huge page management
\cite{windows,ingens,illuminator}. We believe that insights from HawkEye
can be leveraged to improve huge page support in these important systems.
}
\section{Conclusion}

Application transparent support is crucial to the success of large pages. While OS support for 2MB pages has matured over the years, 1GB pages have received little attention despite being present in the hardware for a decade. We propose \trident{} to leverage architectural support of all three page sizes available on x86-64 systems, while also dealing with the latency and fragmentation challenges. Our evaluation shows that transparent 1GB pages, particularly in tandem with 2MB pages, provide a significant performance boost over 2MB THP in Linux. Further, the paravirtualized extension of \trident{}, called \tridentpv{}, can effectively virtualize 1GB pages with copy-less guest page promotion and compaction. We hope that \trident{}'s 18\% performance improvement over Linux THP motivates researchers to further explore the role of large pages, beyond 2MB, in building efficient large-memory systems.
%}

\trident{} source will be available at \url{https://github.com/csl-iisc/Trident}.

\bibliographystyle{plain}
\bibliography{refs.bib}

\begin{thebibliography}{10}

\bibitem{async-zero}
Clearing pages in idle loop.
\newblock
  \url{https://www.mail-archive.com/freebsd-hackers@freebsd.org/msg13993.html"}.

\bibitem{graph500}
Graph500.
\newblock \url{https://graph500.org/}.

\bibitem{gups}
{GUPS: HPCC RandomAccess benchmark}.
\newblock \url{https://github.com/alexandermerritt/gups}.

\bibitem{hugepages-arch}
Hugepages.
\newblock \url{https://wiki.debian.org/Hugepages}.

\bibitem{svm-dataset}
Libsvm data: Classification (binary class).
\newblock
  \url{https://www.csie.ntu.edu.tw/~cjlin/libsvmtools/datasets/binary.html}.

\bibitem{perf}
Linux perf.
\newblock \url{https://en.wikipedia.org/wiki/Perf\_(Linux)}.

\bibitem{buddy}
Page frame allocation via buddy in linux.
\newblock \url{https://wiki.osdev.org/Page\_Frame\_Allocation}.

\bibitem{wikichip:coffeelake}
Wikichip: Intel coffee lake architecture.
\newblock
  \url{https://en.wikichip.org/wiki/intel/microarchitectures/coffee\_lake}.

\bibitem{wikichip:sandybridge}
Wikichip: Intel sandy bridge microarchitecture.
\newblock \url
  {https://en.wikichip.org/wiki/intel/microarchitectures/sandy\_bridge\_(client)}.

\bibitem{mitosis}
Reto Achermann, Ashish Panwar, Abhishek Bhattacharjee, Timothy Roscoe, and
  Jayneel Gandhi.
\newblock Mitosis: Transparently self-replicating page-tables for large-memory
  machines, 2019.

\bibitem{doityourself}
Hanna Alam, Tianhao Zhang, Mattan Erez, and Yoav Etsion.
\newblock Do-it-yourself virtual memory translation.
\newblock In {\em Proceedings of the 44th Annual International Symposium on
  Computer Architecture}, ISCA '17, pages 457--468, New York, NY, USA, 2017.
  ACM.

\bibitem{anandtech:icelake:2019}
AnandTech.
\newblock The ice lake benchmark preview: Inside intel's 10nm, 2019.
\newblock
  \url{https://www.anandtech.com/show/14664/testing-intel-ice-lake-10nm/2}.

\bibitem{nas}
David Bailey, E.~Barszcz, J.~Barton, D.~Browning, Robert Carter, Leonardo
  Dagum, Rod Fatoohi, Paul Frederickson, T.~Lasinski, Robert Schreiber, Horst
  Simon, Venkat Venkatakrishnan, and Sisira Weeratunga.
\newblock The nas parallel benchmarks;summary and preliminary results.
\newblock In {\em Proceedings of the 1991 ACM/IEEE Conference on
  Supercomputing}, Supercomputing '91, pages 158--165, New York, NY, USA, 1991.
  ACM.

\bibitem{pwc}
Thomas~W. Barr, Alan~L. Cox, and Scott Rixner.
\newblock Translation caching: Skip, don't walk (the page table).
\newblock In {\em Proceedings of the 37th Annual International Symposium on
  Computer Architecture}, ISCA '10, pages 48--59, New York, NY, USA, 2010. ACM.

\bibitem{spec-tlb}
Thomas~W. Barr, Alan~L. Cox, and Scott Rixner.
\newblock Spectlb: A mechanism for speculative address translation.
\newblock In {\em Proceedings of the 38th Annual International Symposium on
  Computer Architecture}, ISCA '11, pages 307--318, New York, NY, USA, 2011.
  ACM.

\bibitem{basu:isca:2013}
Arkaprava Basu, Jayneel Gandhi, Jichuan Chang, Mark~D. Hill, and Michael~M.
  Swift.
\newblock Efficient virtual memory for big memory servers.
\newblock In {\em Proceedings of the 40th Annual International Symposium on
  Computer Architecture}, ISCA '13, pages 237--248, New York, NY, USA, 2013.
  ACM.

\bibitem{gapbs}
Scott Beamer, Krste Asanovic, and David~A. Patterson.
\newblock The {GAP} benchmark suite.
\newblock {\em CoRR}, abs/1508.03619, 2015.

\bibitem{2dwalks}
Ravi Bhargava, Benjamin Serebrin, Francesco Spadini, and Srilatha Manne.
\newblock Accelerating two-dimensional page walks for virtualized systems.
\newblock In {\em Proceedings of the 13th International Conference on
  Architectural Support for Programming Languages and Operating Systems},
  ASPLOS XIII, pages 26--35, New York, NY, USA, 2008. ACM.

\bibitem{colt-mmu}
Abhishek Bhattacharjee.
\newblock Large-reach memory management unit caches.
\newblock In {\em Proceedings of the 46th Annual IEEE/ACM International
  Symposium on Microarchitecture}, MICRO-46, pages 383--394, New York, NY, USA,
  2013. ACM.

\bibitem{parsec}
Christian Bienia.
\newblock {\em Benchmarking Modern Multiprocessors}.
\newblock PhD thesis, Princeton University, January 2011.

\bibitem{redisbook}
Josiah~L. Carlson.
\newblock {\em Redis in Action}.
\newblock Manning Publications Co., Greenwich, CT, USA, 2013.

\bibitem{colt-hugepages}
Guilherme Cox and Abhishek Bhattacharjee.
\newblock Efficient address translation for architectures with multiple page
  sizes.
\newblock In {\em Proceedings of the Twenty-Second International Conference on
  Architectural Support for Programming Languages and Operating Systems},
  ASPLOS '17, pages 435--448, New York, NY, USA, 2017. ACM.

\bibitem{intel:3dxpoint}
Ian Cutress.
\newblock {Intel’s Enterprise Extravaganza 2019: Launching Cascade Lake,
  Optane DCPMM, Agilex FPGAs, 100G Ethernet, and Xeon D-1600}.
\newblock
  \url{https://www.anandtech.com/show/14155/intels-enterprise-extravaganza-2019-roundup},
  2019.

\bibitem{superpages-non-contiguous}
Yu~Du, Miao Zhou, Bruce~R. Childers, Daniel Moss{\'e}, and Rami~G. Melhem.
\newblock Supporting superpages in non-contiguous physical memory.
\newblock {\em 2015 IEEE 21st International Symposium on High Performance
  Computer Architecture (HPCA)}, pages 223--234, 2015.

\bibitem{memcached}
Brad Fitzpatrick.
\newblock Distributed caching with memcached.
\newblock {\em Linux J.}, 2004(124):5, August 2004.

\bibitem{carrefour-lp}
Fabien Gaud, Baptiste Lepers, Jeremie Decouchant, Justin Funston, Alexandra
  Fedorova, and Vivien Qu{\'e}ma.
\newblock Large pages may be harmful on numa systems.
\newblock In {\em Proceedings of the 2014 USENIX Conference on USENIX Annual
  Technical Conference}, USENIX ATC'14, pages 231--242, Berkeley, CA, USA,
  2014. USENIX Association.

\bibitem{spgorman-characteristics}
Mel Gorman and Patrick Healy.
\newblock Performance characteristics of explicit superpage support.
\newblock In {\em Proceedings of the 2010 International Conference on Computer
  Architecture}, ISCA'10, pages 293--310, Berlin, Heidelberg, 2012.
  Springer-Verlag.

\bibitem{netflix:thp}
Brendan Gregg.
\newblock How netflix tunes ec2 instances for performance, 2017.
\newblock
  \url{http://www.brendangregg.com/Slides/AWSreInvent2017\_performance\_tuning\_EC2.pdf}.

\bibitem{guvenilir:tps:2020}
Faruk Guvenilir and Yale~N Patt.
\newblock Tailored page sizes.
\newblock In {\em 2020 ACM/IEEE 47th Annual International Symposium on Computer
  Architecture (ISCA)}, pages 900--912. IEEE, 2020.

\bibitem{kwon:osdi:2016}
Youngjin Kwon, Hangchen Yu, Simon Peter, Christopher~J. Rossbach, and Emmett
  Witchel.
\newblock Coordinated and efficient huge page management with ingens.
\newblock In {\em Proceedings of the 12th USENIX Conference on Operating
  Systems Design and Implementation}, OSDI'16, page 705–721, USA, 2016.
  USENIX Association.

\bibitem{compiler-driven}
Joshua Magee and Apan Qasem.
\newblock A case for compiler-driven superpage allocation.
\newblock In {\em Proceedings of the 47th Annual Southeast Regional
  Conference}, ACM-SE 47, pages 82:1--82:4, New York, NY, USA, 2009. ACM.

\bibitem{optane:usecases:2019}
Kristie Mann.
\newblock Five use cases of intel optane dc persistent memory at work in the
  data center, 2019.
\newblock \url{https://itpeernetwork.intel.com/intel-optane-use-cases/}.

\bibitem{asap:micro:2019}
Artemiy Margaritov, Dmitrii Ustiugov, Edouard Bugnion, and Boris Grot.
\newblock Prefetched address translation.
\newblock In {\em Proceedings of the 52nd Annual IEEE/ACM International
  Symposium on Microarchitecture}, MICRO '52, page 1023–1036, New York, NY,
  USA, 2019. Association for Computing Machinery.

\bibitem{navarro}
Juan Navarro, Sitararn Iyer, Peter Druschel, and Alan Cox.
\newblock Practical, transparent operating system support for superpages.
\newblock {\em SIGOPS Oper. Syst. Rev.}, 36(SI):89--104, December 2002.

\bibitem{panwar:asplos:2019}
Ashish Panwar, Sorav Bansal, and K.~Gopinath.
\newblock Hawkeye: Efficient fine-grained os support for huge pages.
\newblock In {\em Proceedings of the Twenty-fourth International Conference on
  Architectural Support for Programming Languages and Operating Systems},
  ASPLOS '19, New York, NY, USA, 2019. ACM.

\bibitem{panwar:asplos:2018}
Ashish Panwar, Aravinda Prasad, and K.~Gopinath.
\newblock Making huge pages actually useful.
\newblock In {\em Proceedings of the Twenty-Third International Conference on
  Architectural Support for Programming Languages and Operating Systems},
  ASPLOS '18, pages 679--692, New York, NY, USA, 2018. ACM.

\bibitem{park:isca:2020}
Chang~Hyun Park, Sanghoon Cha, Bokyeong Kim, Youngjin Kwon, David
  Black-Schaffer, and Huh Jaehyuk.
\newblock Perforated page: Supporting fragmented memory allocation for large
  pages.
\newblock In {\em Proceedings of the 47th International Symposium on Computer
  Architecture}, ISCA '20, New York, NY, USA, 2020. ACM.

\bibitem{colt-hpca}
Binh Pham, Abhishek Bhattacharjee, Yasuko Eckert, and Gabriel~H. Loh.
\newblock Increasing tlb reach by exploiting clustering in page translations.
\newblock {\em 2014 IEEE 20th International Symposium on High Performance
  Computer Architecture (HPCA)}, pages 558--567, 2014.

\bibitem{colt}
Binh Pham, Viswanathan Vaidyanathan, Aamer Jaleel, and Abhishek Bhattacharjee.
\newblock Colt: Coalesced large-reach tlbs.
\newblock In {\em Proceedings of the 2012 45th Annual IEEE/ACM International
  Symposium on Microarchitecture}, MICRO-45, pages 258--269, Washington, DC,
  USA, 2012. IEEE Computer Society.

\bibitem{largepagesvirtual}
Binh Pham, J\'{a}n Vesel\'{y}, Gabriel~H. Loh, and Abhishek Bhattacharjee.
\newblock Large pages and lightweight memory management in virtualized
  environments: Can you have it both ways?
\newblock In {\em Proceedings of the 48th International Symposium on
  Microarchitecture}, MICRO-48, pages 1--12, New York, NY, USA, 2015. ACM.

\bibitem{puttaswamy:gvlsi:2006}
Kiran Puttaswamy and Gabriel Loh.
\newblock Thermal analysis of a 3d die-stacked high-performance microprocessor.
\newblock In {\em Proceedings of the 16th ACM Great Lakes Symposium on VLSI},
  GLSVLSI '06, pages 19--24, New York, NY, USA, 2006. ACM.

\bibitem{pomtlb}
Jee~Ho Ryoo, Nagendra Gulur, Shuang Song, and Lizy~K. John.
\newblock Rethinking tlb designs in virtualized environments: A very large
  part-of-memory tlb.
\newblock In {\em Proceedings of the 44th Annual International Symposium on
  Computer Architecture}, ISCA '17, pages 469--480, New York, NY, USA, 2017.
  ACM.

\bibitem{sodani:micro:2011}
Avinash Sodani.
\newblock Micro keynote: Race to exascale, 2011.
\newblock
  \url{https://www.microarch.org/micro44/files/Micro\%20Keynote\%20Final\%20-\%20Avinash\%20Sodani.pdf}.

\bibitem{xsbench}
John~R Tramm, Andrew~R Siegel, Tanzima Islam, and Martin Schulz.
\newblock {XSBench - The Development and Verification of a Performance
  Abstraction for Monte Carlo Reactor Analysis}.
\newblock In {\em PHYSOR 2014 - The Role of Reactor Physics toward a
  Sustainable Future}, Kyoto.

\bibitem{nvidia:1g:2019}
Zi~Yan.
\newblock Generating physically contiguous memory after page allocation, 2019.
\newblock \url{https://patchwork.kernel.org/cover/10815945/}.

\bibitem{thp-1g:2020}
Zi~Yan.
\newblock 1gb pud thp support on x86\_64.
\newblock \url{https://lwn.net/Articles/832881/}, 2020.

\bibitem{yan:isca:2019}
Zi~Yan, Daniel Lustig, David Nellans, and Abhishek Bhattacharjee.
\newblock Translation ranger: Operating system support for contiguity-aware
  tlbs.
\newblock In {\em Proceedings of the 46th International Symposium on Computer
  Architecture}, ISCA '19, pages 698--710, New York, NY, USA, 2019. ACM.

\bibitem{quicksilver:atc:2020}
Weixi Zhu, Alan~L. Cox, and Scott Rixner.
\newblock A comprehensive analysis of superpage management mechanisms and
  policies.
\newblock In {\em 2020 {USENIX} Annual Technical Conference ({USENIX} {ATC}
  20)}, pages 829--842. {USENIX} Association, July 2020.

\end{thebibliography}

\end{document}